\documentclass[aps,prd,reprint,preprintnumbers,superscriptaddress,showpacs,twocolumn]{revtex4-1}
\usepackage{latexsym,graphicx,amssymb,amsmath,mathrsfs}
\usepackage{setspace,bm}
\usepackage[breaklinks, colorlinks=true, pdfstartview=FitV, linkcolor=red, citecolor=blue, urlcolor=blue]{hyperref}
\usepackage[usenames]{color}
\usepackage{latexsym}
\usepackage{epstopdf}
\usepackage{mathtools}
\usepackage{url}
\usepackage{comment}
\usepackage{braket}
\usepackage{CJKutf8}

\newcommand{\bequ}{\begin{equation}}
\newcommand{\eequ}{\end{equation}}
\newcommand{\bea}{\begin{eqnarray}}
\newcommand{\eea}{\end{eqnarray}}

\DeclareSymbolFont{boldletters}{OML}{cmm} {b}{it}
\DeclareSymbolFontAlphabet{\mathbit}{boldletters}
\DeclareMathSymbol{\alpha}{\mathalpha}{letters}{"0B}
\DeclareMathSymbol{\beta}{\mathalpha}{letters}{"0C}
\DeclareMathSymbol{\gamma}{\mathalpha}{letters}{"0D}
\DeclareMathSymbol{\delta}{\mathalpha}{letters}{"0E}
\DeclareMathSymbol{\epsilon}{\mathalpha}{letters}{"0F}
\DeclareMathSymbol{\zeta}{\mathalpha}{letters}{"10}
\DeclareMathSymbol{\eta}{\mathalpha}{letters}{"11}
\DeclareMathSymbol{\theta}{\mathalpha}{letters}{"12}
\DeclareMathSymbol{\iota}{\mathalpha}{letters}{"13}
\DeclareMathSymbol{\kappa}{\mathalpha}{letters}{"14}
\DeclareMathSymbol{\lambda}{\mathalpha}{letters}{"15}
\DeclareMathSymbol{\mu}{\mathalpha}{letters}{"16}
\DeclareMathSymbol{\nu}{\mathalpha}{letters}{"17}
\DeclareMathSymbol{\xi}{\mathalpha}{letters}{"18}
\DeclareMathSymbol{\pi}{\mathalpha}{letters}{"19}
\DeclareMathSymbol{\rho}{\mathalpha}{letters}{"1A}
\DeclareMathSymbol{\sigma}{\mathalpha}{letters}{"1B}
\DeclareMathSymbol{\tau}{\mathalpha}{letters}{"1C}
\DeclareMathSymbol{\upsilon}{\mathalpha}{letters}{"1D}
\DeclareMathSymbol{\phi}{\mathalpha}{letters}{"1E}
\DeclareMathSymbol{\chi}{\mathalpha}{letters}{"1F}
\DeclareMathSymbol{\psi}{\mathalpha}{letters}{"20}
\DeclareMathSymbol{\omega}{\mathalpha}{letters}{"21}
\DeclareMathSymbol{\varepsilon}{\mathalpha}{letters}{"22}
\DeclareMathSymbol{\vartheta}{\mathalpha}{letters}{"23}
\DeclareMathSymbol{\varpi}{\mathalpha}{letters}{"24}
\DeclareMathSymbol{\varrho}{\mathalpha}{letters}{"25}
\DeclareMathSymbol{\varsigma}{\mathalpha}{letters}{"26}
\DeclareMathSymbol{\varphi}{\mathalpha}{letters}{"27}
\DeclareMathSymbol{\Gamma}{\mathalpha}{letters}{"00}
\DeclareMathSymbol{\Delta}{\mathalpha}{letters}{"01}
\DeclareMathSymbol{\Theta}{\mathalpha}{letters}{"02}
\DeclareMathSymbol{\Lambda}{\mathalpha}{letters}{"03}
\DeclareMathSymbol{\Xi}{\mathalpha}{letters}{"04}
\DeclareMathSymbol{\Pi}{\mathalpha}{letters}{"05}
\DeclareMathSymbol{\Sigma}{\mathalpha}{letters}{"06}
\DeclareMathSymbol{\Upsilon}{\mathalpha}{letters}{"07}
\DeclareMathSymbol{\Phi}{\mathalpha}{letters}{"08}
\DeclareMathSymbol{\Psi}{\mathalpha}{letters}{"09}
\DeclareMathSymbol{\Omega}{\mathalpha}{letters}{"0A}

\begin{document}
\title{Roberge-Weiss periodicity and singularity in hadron resonance gas model with excluded volume effects}

\author{Riki Oshima}
\email[]{24804001@edu.cc.saga-u.ac.jp}
\affiliation{Department of Physics, Saga University,
             Saga 840-8502, Japan}

\author{Hiroaki Kouno}
\email[]{kounoh@cc.saga-u.ac.jp}
\affiliation{Department of Physics, Saga University,
             Saga 840-8502, Japan}

\author{Kouji Kashiwa}
\email[]{kashiwa@fit.ac.jp}
\affiliation{Fukuoka Institute of Technology, Wajiro, Fukuoka 811-0295, Japan}


\begin{abstract}
Quantum chromodynamics (QCD) with pure imaginary baryon number chemical potential $\mu =i\theta T$, where $T$ is temperature and $\theta$ is a real number, has the Roberge-Weiss periodicity. 
We study the $\theta$-dependence of the baryon number density and the pressure in the hadron resonance gas model with excluded volume effects of baryons. 
It is shown that the baryon number density and the pressure are smooth periodic functions of $\theta$ at low or high temperature.  
However, they have singular behavior at $\theta =(2k+1)\pi$ where $k$ is an integer, when $T\sim 211$~MeV.  
This temperature is consistent with the Roberge-Weiss transition temperature $T_{\rm RW}$ obtained by lattice QCD simulations.
This singularity can be explained by the dual excluded volume effects in which the roles of point-like and non point-like particles are exchanged each other in the ordinary excluded volume effects.  
It is also indicated that the excluded volume effect is visible just below $T_{\rm RW}$ and is directly detectable by the lattice QCD simulation at finite $\theta$. 
We compare the results with the one obtained by the Polyakov-loop extended Nambu--Jona-Lasinio model.
\end{abstract}

\maketitle


\section{Introduction}

Determination of the phase diagram of quantum chromodynamics (QCD) is an important subject not only in nuclear and particle physics but also in cosmology and astrophysics; see, e.g., Ref.~\cite{Fukushima:2010bq} and references therein.  
However, when the baryon number chemical potential is finite and real, the fist principle calculation, namely, the lattice QCD (LQCD) simulation, is not feasible due to the infamous sign problem; see Refs.\,\cite{deForcrand:2010ys,Nagata:2021bru,*Nagata:2021ugx} as an example. 
To circumvent the sign problem, several methods are proposed and investigated, although, at present, these methods are not complete and we do not have adequate information on the equation of state (EoS) at finite real baryon density. 
 
One possible way to avoid the sign problem is to use the LQCD results with the imaginary baryon number chemical potential; see Refs.\,\cite{deForcrand:2002ci,D'Elia:2002gd,D'Elia:2004at,Chen:2004tb,D'Elia:2009qz} as an example.
When the baryon number chemical potential $\mu$ is pure imaginary, there is no sign problem. 
One can perform LQCD simulations at finite pure imaginary $\mu$, and then make an analytic continuation from the quantities at imaginary $\mu$ to those at real $\mu$. 
Alternatively, one may determine the unknown parameters of an effective model of QCD using the LQCD results at finite imaginary $\mu$. 
After determining the parameters, the model calculations can be performed at real $\mu$~\cite{Sakai:2009dv}.  
  
It is known that the LQCD results at $\mu = 0$ are in good agreement with those obtained by the hadron resonance gas (HRG) model when temperature $T$ is not so large.  
Usually, the ideal gas approximation is used for the calculations in the HRG model.  
However, it is expected that repulsive effects among baryons are important at high baryon number density. 
If the repulsion is absent, baryon matter is realized at a sufficiently large baryon density~\cite{Cleymans:1985wb}. 
One of the traditional treatments for such repulsion is to consider excluded volume effects (EVE) among baryons~\cite{Cleymans:1986cq, Kouno:1988bi, Rischke:1991ke,Miyahara:2019zfn,Jeong:2019lhv}.  
EVE successfully prevents baryon matter from realizing at sufficiently large baryon density~\cite{Cleymans:1986cq}; for the recent review, see, e.g., Ref.~\cite{Fujimoto:2021dvn} and references therein.
The availability of the HRG model with EVE may be checked by using the LQCD results at finite imaginary $\mu$. 

The grand canonical QCD partition function $Z(\theta)$ with pure imaginary quark chemical potential ($\mu_{\rm q}={\mu / 3}=i\theta_{\rm q} T$) has the Roberge-Weiss (RW) periodicity~\cite{Roberge:1986mm} as
\begin{eqnarray}
Z \Bigl(\theta_{\rm q} + 
       {2\pi\over{3}} \Bigr) 
= Z(\theta_{\rm q}), 
\label{RWP1}
\end{eqnarray}
where $T$ is the temperature and $\theta_\mathrm{q} \in \mathbb{R}$. 
This periodicity is the remnant of the $\mathbb{Z}_3$-symmetry of pure gluon theory. 
At low temperature, $Z(\theta_{\rm q} )$ is expected to be a smooth function of $\theta_{\rm q}$. 
However, at high temperature above the RW temperature $T_{\rm RW}$, it has a singularity at $\theta_{\rm q}=(2k+1) \pi / 3$ where $k \in \mathbb{Z}$.   
This singularity is called the RW transition. 
$T_{\rm RW}$ for 2+1 flavor QCD is estimated as about 200~MeV by LQCD simulations~\cite{Bonati:2016pwz,Cuteri:2022vwk,Bonati:2018fvg}. 

In the HRG model with pure imaginary baryon number chemical potential $\mu =i\theta T$, the RW periodicity is trivial, since the model has a trivial periodicity
\begin{eqnarray}
Z_{\rm HRG}(\theta  +2\pi )=Z_{\rm HRG}(\theta), 
\label{RWP_HRG}
\end{eqnarray}
and $\theta :=3\theta_q$.  
In the case of the free hadron resonance gas, $Z_{\rm HRG}(\theta )$ is a smooth function of $\theta$ at any temperature. 
However, it may have a singularity when interaction effects such as EVE are taken into account. 

In this paper, we study the $\theta$-dependence of the baryon number density and the pressure in the HRG model with EVE when the imaginary baryon number chemical potential is introduced, and analyze the mechanism of EVE at finite imaginary $\mu$. 
It is shown that the baryon number density and the pressure have a singular behavior at $\theta =(2k+1)\pi$ when $T\sim 211$~MeV, while they are smooth functions of $\theta$ at lower and higher temperature. 
This result is consist with the previous result in Refs.~\cite{Taradiy:2019taz,Savchuk:2019yxl} where the singularity of the HRG model with EVE was reported and its impacts were discussed, although our model is somewhat different from the previous ones. 
The singularity is understood by the dual EVE in which the roles of point-like and non point-like particles are exchanged. 
The temperature $T\sim 211$~MeV is compatible with $T_{\rm RW}$ estimated by LQCD simulations. 
We also compare the results with the ones obtained by the Polyakov-loop extended Nambu--Jona-Lasinio (PNJL) model~\cite{Fukushima:2003fw,Ratti:2005jh,Ghosh:2006qh,Megias:2004hj,Roessner:2006xn,Sakai:2010rp} in which the quark degree of freedom is contained. 

This paper is organized as follows. 
In Sec.~\ref{RWP}, the RW periodicity and transition are briefly reviewed.  
In Sec.~\ref{formalism}, we show our formulation of the HRG model with EVE. 
The concept of dual EVE is also explained. 
In Sec. \ref{Nresults}, numerical results are shown. 
Section \ref{summary} is devoted to the summary and discussions.

\section{Roberge-Weiss periodicity and transition}
\label{RWP}

The grand canonical partition function of QCD with imaginary $\mu =i\theta_{\rm q}T$ is given by
\begin{align}
Z(\theta_{\rm q})
&= \int {\cal D} \psi {\cal D} \bar{\psi} {\cal D} A_\mu e^{-S(\theta_{\rm q} )}, 
\label{Z-QCD}
\end{align}
where
\begin{align}
S(\theta_{\rm q})
&=\int_0^\beta d\tau\int_{-\infty}^\infty d^3x \, {\cal L}(\theta_{\rm q}),
\label{S-QCD}
\end{align}
with
\begin{align}
{\cal L} (\theta_{\rm q})
&= \bar{\psi}(\gamma_\mu D_\mu -m_0)\psi 
 -{1\over{4}}F_{\mu\nu}^2-i{\theta_{\rm q}\over{\beta}}\bar{\psi}\gamma_4\psi ,
\label{L-QCD}
\end{align}
here $\psi$, $A_\mu$, $D_\mu$, $F_{\mu\nu}$ and $m_0$ are the quark field, the gluon field, the covariant derivative, the field strength of gluon and the current quark mass matrix, respectively, and $\beta = 1/T$. 
The Euclidean notation is used in Eqs. (\ref{Z-QCD}) $\sim$ (\ref{L-QCD}).  

To eliminate the $\theta_{\rm q}$-term from the action, we perform the following transformation of quark field; 
\begin{eqnarray}
\psi \mapsto \exp{\left(i{\tau \theta_{\rm q}\over{\beta}}\right)}\psi. 
\label{transform}
\end{eqnarray}
However, as a result, the anti-periodic temporal boundary condition $\psi$ is changed as follows;
\begin{eqnarray}
\psi ({\bf x},\beta )=-\exp{(i\theta_{\rm q})}\psi ({\bf x},0). 
\label{quark_boundary_1}
\end{eqnarray}
This means that $\theta_{\rm q}$ can be considered as the phase of the temporal boundary condition of quarks. 

We perform another change of the quark and gluon fields;  
\begin{eqnarray}
A_\mu&\mapsto& U({\bf x},\tau )A_\mu U^{-1}({\bf x},\tau )-{i\over{g}}(\partial_\mu U({\bf x},\tau ))U^{-1}({\bf x},\tau ), 
\nonumber\\
\psi &\mapsto & U({\bf x},\tau )\psi, 
\label{Z3trans}
\end{eqnarray}
where $g$ is a coupling constant, $U({\bf x},\tau)$ are elements of $SU(3)$ with the temporal boundary condition $U({\bf x},\beta)=z_3U({\bf x},0)$ and 
$z_3=\exp(i2\pi k/3)$ is a $\mathbb{Z}_3$ element. 
The action $S(\theta )$ is invariant under this ${\mathbb Z}_3$  transformation but the quark boundary condition is changed as 
\begin{eqnarray}
\psi ({\bf x},\beta )=-\exp{\Big[i\Big( \theta_{\rm q} +{2\pi k\over{3}}\Big)\Big]}\psi ({\bf x},0). 
\label{quark_boundary}
\end{eqnarray}
Hence, under the $\mathbb{Z}_3$ transformation, $\theta_{\rm q}$ in $Z(\theta_{\rm q})$ is changed into $\theta_{\rm q} +{2\pi k\over{3}}$ and we obtain the RW periodicity (\ref{RWP1}) ~\cite{Roberge:1986mm}. 
Dynamical quarks break the $\mathbb{Z}_3$-symmetry but the RW periodicity appears as a remnant of the $\mathbb{Z}_3$-symmetry. 

\begin{figure}[t]
\centering
\includegraphics[width=0.40\textwidth]{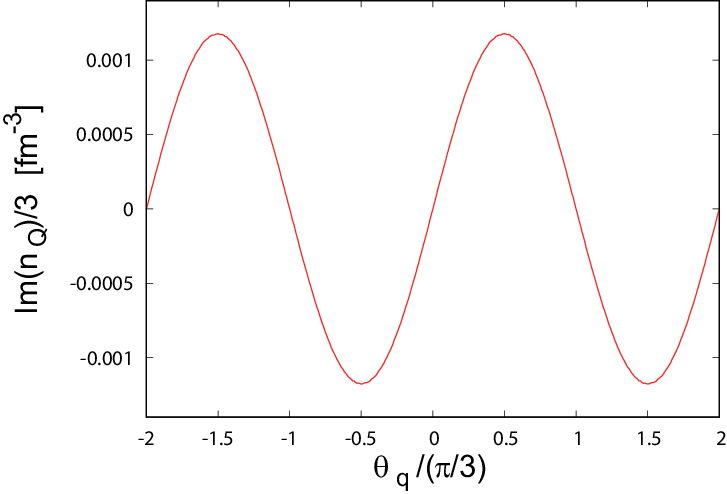}\\
\bigskip
~\includegraphics[width=0.39\textwidth]{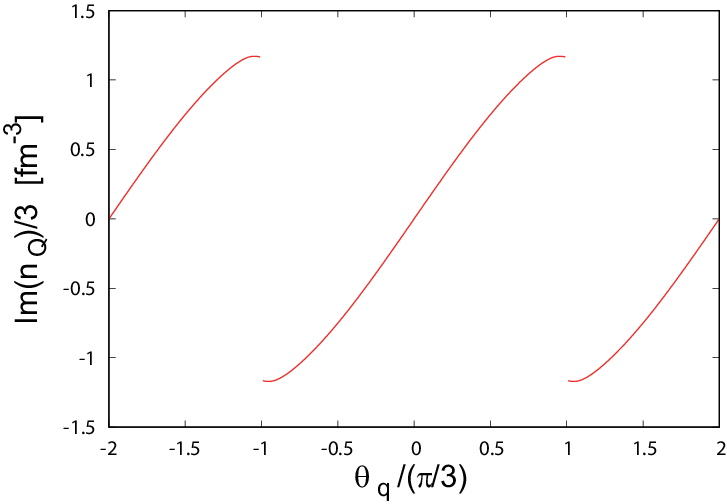}\\
\caption{The top (bottom) panel shows the $\theta_{\rm q}$-dependence of the imaginary part of the baryon number density $n_{\rm q} / 3$ when $T=150$ MeV ($250$ MeV) in the PNJL model; see Appendix~\ref{PNJLmodel} for details of the model. 
Note that $T_{\rm RW}=201$~MeV in this model. 
The discontinuities of $n_{\rm q}$ appear at $\theta_{\rm q}=\pm\pi/3$ in the bottom panel. 
}
\label{Fig_T=150_250_PNJL}
\end{figure}

\begin{figure}[t]
\centering
\includegraphics[width=0.40\textwidth]{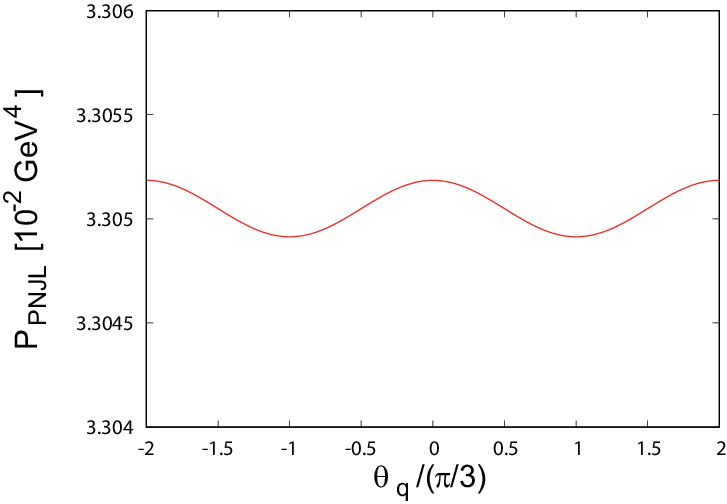}\\
\bigskip
~\includegraphics[width=0.39\textwidth]{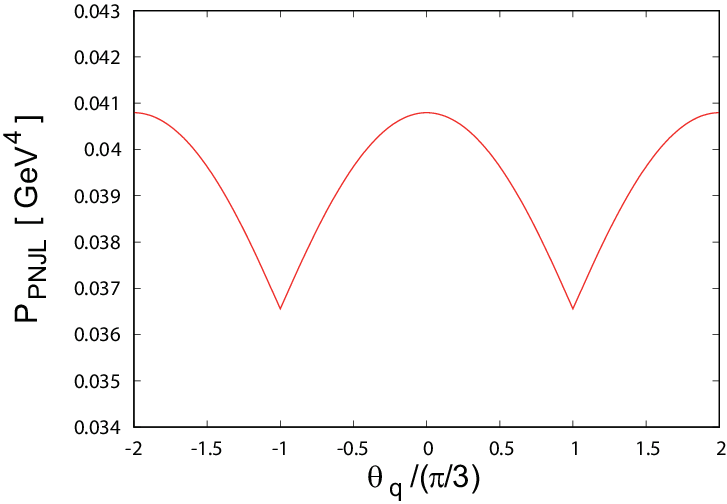}\\
\caption{The top (bottom) panel shows the $\theta_{\rm q}$-dependence of the pressure $P_{\rm PNJL}$ when $T=150$ MeV ($250$ MeV) in the PNJL model; see 
The cusps of $P_{\rm PNJL}$ appear at $\theta_{\rm q}=\pm\pi/3$ in the bottom panel. 
}
\label{Fig_PNJL_P}
\end{figure}

At low temperature below the RW transition temperature $T_{\rm RW}$, the baryon number density $n_{\rm Q}/ 3$, where $n_{\rm Q}$ is the quark number density, is a smooth function of $\theta$.  
However, at high temperature above $T_{\rm RW}$, it is discontinuous at $\theta_{\rm q}=(2k+1)\pi/3$ due to the degeneracy of the ground state.  
For illustrations, in Fig.\,\ref{Fig_T=150_250_PNJL},
we show the $\theta_{\rm q}$-dependence of $n_{\rm Q}/3$ obtained by the PNJL  model~\cite{Fukushima:2003fw,Ratti:2005jh,Ghosh:2006qh,Megias:2004hj,Roessner:2006xn,Sakai:2010rp} which is one of the most successful effective models of QCD. 
It is well known that the PNJL model can reproduce several important features of QCD at finite imaginary $\mu$; for example, see Ref.\,\cite{Kashiwa:2019ihm} as a review. 
We also show the $\theta_{\rm q}$-dependence of pressure $P_{\rm PNJL}$ in Fig.~\ref{Fig_PNJL_P}. $P_{\rm PNJL}$ is a smooth function of $\theta_{\rm q}$ when $T<T_{\rm RW}$, while it has a cusp at $\theta_{\rm q}=\pm \pi/3$ when $T>T_{\rm RW}$.   
The RW periodicity and the RW transition are also confirmed by LQCD simulations
and $T_{\rm RW}$ is estimated as $195\sim 208$~MeV for the 2+1 flavor LQCD simulation~\cite{Bonati:2016pwz,Cuteri:2022vwk,Bonati:2018fvg}.

\section{Formulation}
\label{formalism}

In this section, we show the detailed formulation of the HRG model with EVE at finite $T$. 

\subsection{Excluded volume effects}

\begin{figure}[t]
\centering
\includegraphics[width=0.3\textwidth]{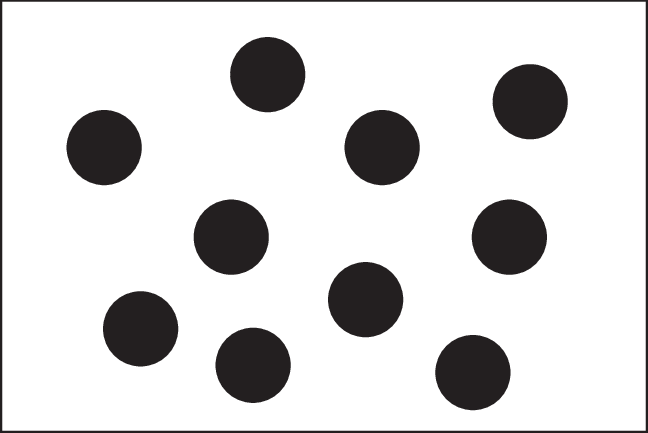}\\
\bigskip
\includegraphics[width=0.3\textwidth]{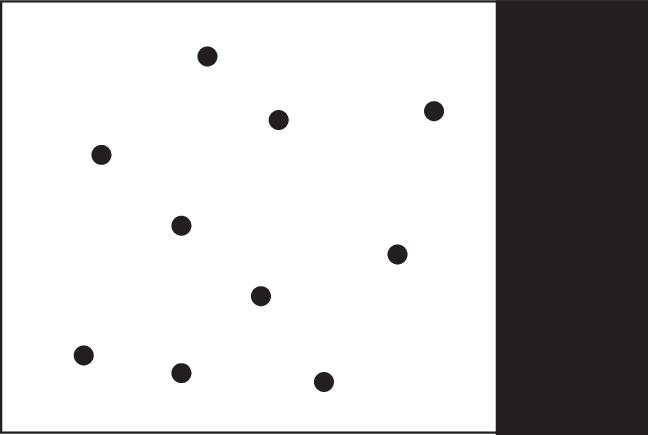}\\
\caption{The top and bottom panels show the schematic figure of the $N$ non point-like particles in volume $V$ and
the $N$ point-like particles in the effective volume $V-vN$, respectively. 
}
\label{Fig_EVE_1}
\end{figure}

Consider $N$ non point-like particles in the system with volume $V$; see the top panel of Fig.~\ref{Fig_EVE_1}.   
We regard that this system is equivalent to the system of $N$ point-like particle in the effective volume $V-v\,N$ where $v$ is the volume of a non point-like particle~\cite{Cleymans:1986cq}; see the bottom panel of Fig.~\ref{Fig_EVE_1}. 
Hence, we obtain 
\begin{eqnarray}
n_{\rm p}={N\over{V-v\,N}}={n\over{1-v\,n}}, 
\label{EVE_1} 
\end{eqnarray}
where $n=N / V$ is the number density of non point-like particles and $n_{\rm p}$ is the one of $N$ point-like particles. 
Then, we obtain the number density $n$ as
\begin{eqnarray}
n={n_{\rm p}\over{1+v\,n_{\rm p}}}.    
\label{EVE_2} 
\end{eqnarray}
Note that Eqs.~(\ref{EVE_1}) and (\ref{EVE_2}) are valid only when the denominator is not zero. 
$n$ can not exceed the upper bound $1 / v$ when $n_{\rm p}$ is real and positive, and $n \to 1/v$ when $n_{\rm p} \to \infty$. 

If we replace $v$ by $-v$, Eq.~(\ref{EVE_1}) is changed into
\begin{eqnarray}
n_{\rm p}={n\over{1+v\,n}}. 
\label{EVE_1_dual} 
\end{eqnarray}
Hence, $n_{\rm p}\to 1/v$, when $n\to \infty$. 
In this paper, we call this effect "dual EVE". 
The dual EVE seems to be unnatural, since it is equivalent to EVE with a negative particle volume. 
However, later, we show that dual EVE actually occurs when the chemical potential is pure imaginary.

\subsection{Hadron resonance gas model with excluded volume effects at finite temperature}

Here, we consider the HRG model with EVE. 
At zero temperature, EVE is usually introduced for the baryon number $N_{\rm B}$. 
However, at finite temperature, the number $N_{\rm b}$ of baryons and the number $N_{\rm a}$ of anti-baryons do not vanish even if $N_{\rm B}=N_{\rm b}-N_{\rm a}$ is zero.
Hence, according to the next policy~\cite{Kouno:1988bi}, we introduce EVE into the HRG model.   

~

\noindent
(1) Since EVE represents the effects of the repulsion forces among baryons, a baryon is affected by EVE of other baryons, but is not affected by EVE of anti-baryons.

~

\noindent
(2) Inversely, an anti-baryon is affected by EVE of other anti-baryons, but is not affected by EVE of baryons. 

~

\noindent
(3) Mesons are not affected by EVE. 
We treat the meson gas as a free gas of point-like particles for simplicity.

~

\noindent
Therefore, the baryon number density $n_{\rm B}$ of the system is given by 
\begin{align}
n_{\rm B}(\mu )
&= n_{\rm b}(\mu )-n_{\rm a}(\mu ),
\label{HRG_nB}
\end{align}
with
\begin{align}
n_{\rm b}(\mu )
&= {n_{\rm bp}(\mu )\over{1+v\,n_{\rm bp}(\mu )}}, 
\label{HRG_nb}
\\
n_{\rm a}(\mu )
&= {n_{\rm ap}(\mu )\over{1+v\,n_{\rm ap}(\mu )}}={n_{\rm bp}(-\mu )\over{1+v\,n_{\rm bp}(-\mu )}}=n_{\rm b}(-\mu ), 
\label{HRG_na}
\end{align}
where 
$n_{\rm b}$, $n_{\rm a}$, $n_{\rm bp}$ and $n_{\rm ap}$ is the number densities of baryons, anti-baryons, point-like baryons and point-like anti-baryons, respectively. 
Here, we have assumed that $v$ is the same for all baryons and anti-baryons for simplicity. 
In numerical calculations, we set $v = 4 \pi r_0^3 / 3$ with $r_0=0.8$~fm. 

Note that $n_{\rm B}$ is an odd function of $\mu$. 
Hence, when $\mu$ is a pure imaginary number, $n_{\rm B}$ is also a pure imaginary number, although $n_{\rm b}$ and $n_{\rm a}$ are complex numbers in general. 
The relations 
\begin{align}
{\rm Re}\,n_{\rm b} &= {\rm Re}\,n_{\rm a},
\label{Ren}
\\
{\rm Im}\,n_{\rm b} &= -{\rm Im}\,n_{\rm a}
\label{Imn}
\end{align}
with
\begin{align}
{\rm Re}\,n_{\rm b}
&= {{\rm Re}\,n_{\rm bp}(1+v\,{\rm Re}\,n_{\rm bp})+v({\rm Im}\,n_{\rm bp})^2\over
{(1+v\,{\rm Re}\,n_{\rm bp})^2+v^2({\rm Im}\,n_{\rm bp})^2}},
\nonumber\\
{\rm Im}\,n_{\rm b}
&= {i{\rm Im}\,n_{\rm bp}\over{(1+v\,{\rm Re}\,n_{\rm bp})^2+v^2({\rm Im}\,n_{\rm bp})^2}}, 
\nonumber\\
{\rm Re}\,n_{\rm a}
&= {{\rm Re}\,n_{\rm ap}(1+v\,{\rm Re}\,n_{\rm ap})+v({\rm Im}\,n_{\rm ap})^2\over
{(1+v\,{\rm Re}\,n_{\rm ap})^2+v^2({\rm Im}\,n_{\rm ap})^2}},
\nonumber\\
{\rm Im}\,n_{\rm a}
&= {i{\rm Im}\,n_{\rm ap}\over{(1+v\,{\rm Re}\,n_{\rm ap})^2+v^2({\rm Im}\,n_{\rm ap})^2}}, 
\label{n_form}
\end{align}
are realized since ${\rm Re}\,n_{\rm bp}={\rm Re}\,n_{\rm ap}$ and ${\rm Im}\,n_{\rm bp}=-{\rm Im}\,n_{\rm ap}$, and thus the relation
\begin{align}
n_{\rm B}
&= 2i\,{\rm Im}\,n_{\rm b}=-2i\,{\rm Im}\,n_{\rm a},
\label{nBim}
\end{align}
is manifested for Eq.\,(\ref{HRG_nB}). 

When the free gas approximation is used, the concrete forms of $n_{\rm bp}$ and $n_{\rm ap}$ of $i$-th baryon are given by 
\begin{eqnarray}
n_{{\rm bp}, i}
&=&{g_{{\rm s},i}\over{2\pi^2}}\int_0^\infty~dp~p^2{1\over{\exp{\{\beta (\sqrt{p^2+M_i^2}-\mu)\}}+1}}, 
\nonumber\\
\label{nb_free_point_gas}
\\
n_{{\rm ap},i}
&=&{g_{{\rm s},i}\over{2\pi^2}}\int_0^\infty~dp~p^2 
{1\over{\exp{\{\beta (\sqrt{p^2+M_i^2}+\mu)\}}+1}}, 
\nonumber\\
\label{na_free_point_gas}
\end{eqnarray}
where $g_{{\rm s},i}$ and $M_i$ are  the spin degeneracy and the mass of the $i$-th baryon, respectively. 
If we use the Boltzmann distribution instead of the Fermi distribution, Eqs.~(\ref{nb_free_point_gas}) and (\ref{na_free_point_gas}) are reduced to 
\begin{align}
n_{{\rm bp},i}
&= A_i(\cos{\theta}+i\sin{\theta}), 
\label{nb_Boltzmann}
\\
n_{{\rm ap},i}
&= A_i(\cos{\theta}-i\sin{\theta}),
\label{na_Boltzmann}
\end{align}
with
\begin{align}
A_i &= {g_{{\rm s},i}\over{2\pi^2}}\int_0^\infty~dp~p^2e^{-\beta\sqrt{p^2+M_i^2}},
\label{amplitude}
\end{align}
when $\mu$ is pure imaginary. 
Hence, the relation
\begin{eqnarray}
{\rm Re}\Big[n_{\rm bp}(\theta)\Big]&=&{\rm Re}\Big[n_{\rm ap}(\theta)\Big]
\nonumber\\
&=&{\rm Im}\Big[n_{\rm bp}(\theta +{\pi\over{2}})\Big]
\nonumber\\
&=&-{\rm Im}\Big[n_{\rm ap}(\theta +{\pi\over{2}})\Big], 
\label{np_Re_Im} 
\end{eqnarray}
is satisfied. 


In Ref.\,\cite{Kouno:1988bi}, the following simple form of pressure $P_{\rm B}$ of the baryon and antibaryon system with EVE
was used;
\begin{eqnarray}
P_{\rm B}={P_{\rm bp}\over{1+vn_{\rm bp}}}+{P_{\rm ap}\over{1+vn_{\rm ap}}}, 
\label{PB_old}
\end{eqnarray}
where $P_{\rm bp}$ and $P_{\rm ap}$ are the pressures of point-like baryons and antibaryons, respectively. 
However, this form of pressure does not satisfy the thermodynamical relation ${\partial P_{\rm B}\over{\partial \mu}}=n_{\rm B}$ accurately. 
Then, we do not use the form  (\ref{PB_old}). 
In this paper, 
the $\mu$-dependence of baryon pressure $P_{\rm B}$ with EVE is given by 
\begin{align}
P_{\rm B}(\mu ) &= P_{\rm b}(\mu )+P_{\rm a}(\mu )=P_{\rm b}(\mu )+P_{\rm b}(-\mu ), 
\label{PBaryon}
\end{align}
with
\begin{align}
P_{\rm b}(\mu ) &= \int_0^\mu d\mu^\prime n_{\rm b}(\mu^\prime )+P_{\rm b}(0),
\label{Pb}
\end{align}
where the initial conditions are given by
\begin{eqnarray}
P_{\rm b}(0)&=&\int_{-\infty}^0 d\mu^\prime n_{\rm b}(\mu^\prime ), 
\label{initial_Pb}
\end{eqnarray}
respectively. 
In this formulation, the thermodynamical relation is automatically satisfied. 
Note that it is natural to assume $P_{\rm b}(-\infty)=P_{\rm a}(\infty )=0$. 
In numerical calculations, we include all hadrons that are expected to be composed of a light quark only and are listed in the list of particle data~\cite{Workman:2022ynf}. 

In Refs.~\cite{Taradiy:2019taz,Savchuk:2019yxl,Vovchenko:2017xad}, the different formulation of EVE effects was used. 
In the formulation, the transcendental equations
\begin{eqnarray}
P_\mathrm{b}(T,\mu )&=&P_\mathrm{bp}(T,\mu -b P_\mathrm{b}), 
\label{TEb}
\\
P_\mathrm{a}(T,\mu )&=&P_\mathrm{ap}(T,\mu -b P_\mathrm{a}), 
\label{TEa}
\end{eqnarray}
are required, 
where $P_\mathrm{bp}$($P_\mathrm{ap}$
) is  the pressure of the point-like baryons (antibaryons) and $b$ is the parameter which represents baryon (antibaryon) volume. 
In this paper, we call this formulation EVETC model, and call our formulation EVE model.  
In the next section, we compare the results of our formulation with those of EVETC.

\section{Numerical results}
\label{Nresults}

In this section, we show our numerical results for the HRG model with EVE.
First, we compare the results of the two formulations of EVE. 
Next, we show the $\theta$-dependence of the baryon number density and the pressure, and discuss the effects of EVE below $T_\mathrm{RW}$.
We also show the same quantities above $T_\mathrm{RW}$. 
Finally, we show the $T$-dependence of the pressure at $\theta =\pi$ and $\mu =0~(\theta =0)$. 

\subsection{Comparison of two formulation}

\begin{figure}[h]
\centering
\centerline{\includegraphics[width=0.40\textwidth]{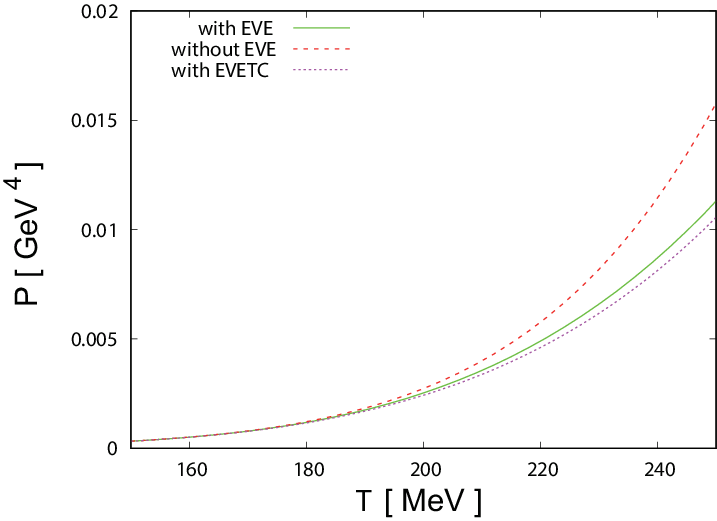}}
\caption{The $T$-dependence of the pressure when $\mu =0$ and $b=v=4\pi r_0^3/3$. 
The solid, dashed, and dotted lines show the results obtained by the HRG models with EVE ($P$), without EVE ($P_{\rm p}$), with EVETC, respectively. 
}
 \label{Fig_P_mu=0_comp}
\end{figure}

\begin{figure}[h]
\centering
\centerline{\includegraphics[width=0.40\textwidth]{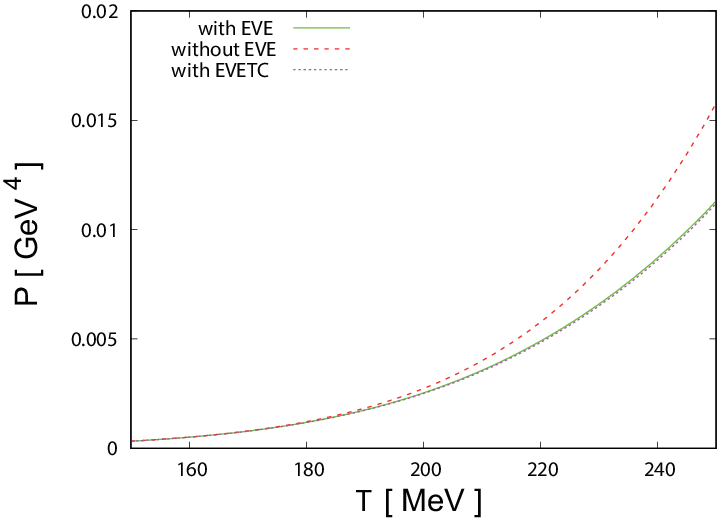}}
\caption{The $T$-dependence of the pressure when $\mu =0$, $v=4\pi r_0^3/3$ and $b=1.315\,\mathrm{fm}^3$. 
The solid, dashed, and dotted lines show the results obtained by the HRG models with EVE ($P$), without EVE ($P_{\rm p}$), with EVETC, respectively. 
}
 \label{Fig_P_mu=0_comp_2}
\end{figure}

\begin{figure}[h]
\centering
\centerline{\includegraphics[width=0.40\textwidth]{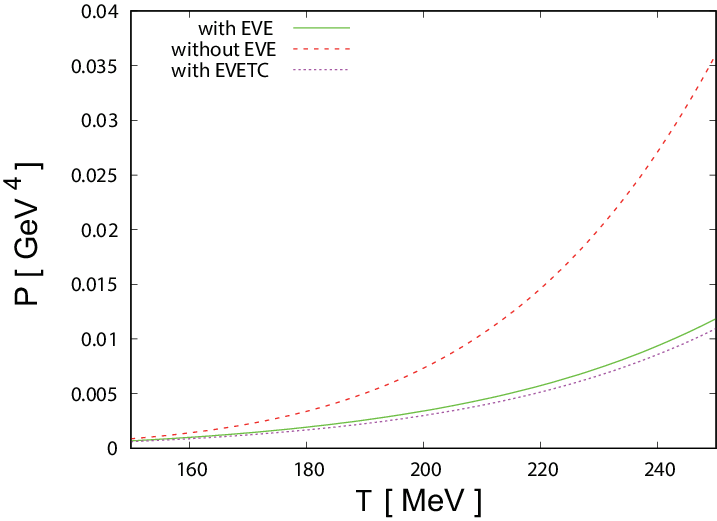}}
\caption{The $T$-dependence of the pressure when $\mu=500$~MeV and $b=v=4\pi r_0^3/3$. 
The solid, dashed, and dotted lines show the results obtained by the HRG models with EVE ($P$), without EVE ($P_{\rm p}$), with EVETC, respectively. 
}
 \label{Fig_P_mu=500_comp}
\end{figure}

\begin{figure}[h]
\centering
\centerline{\includegraphics[width=0.40\textwidth]{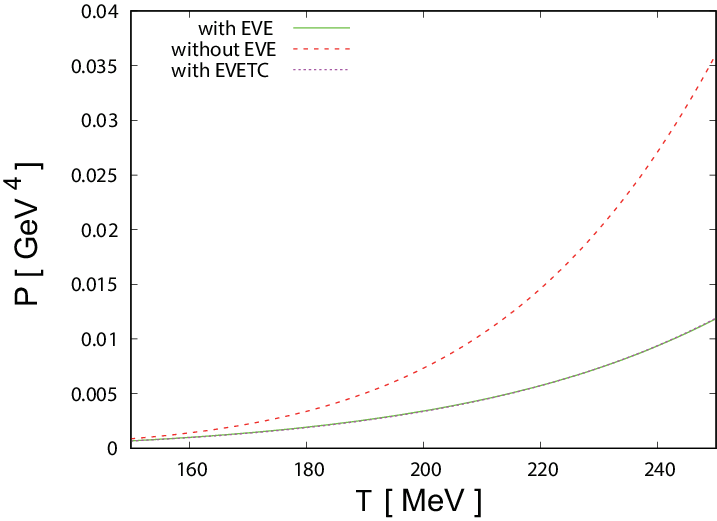}}
\caption{The $T$-dependence of the pressure when $\mu=500$~MeV, $v=4\pi r_0^3/3$ and $b=1.315\,\mathrm{fm}^3$. 
The solid, dashed, and dotted lines show the results obtained by the HRG models with EVE ($P$), without EVE ($P_{\rm p}$), with EVETC, respectively. 
}
 \label{Fig_P_mu=500_comp_2}
\end{figure}

\begin{figure}[h]
\centering
\centerline{\includegraphics[width=0.40\textwidth]{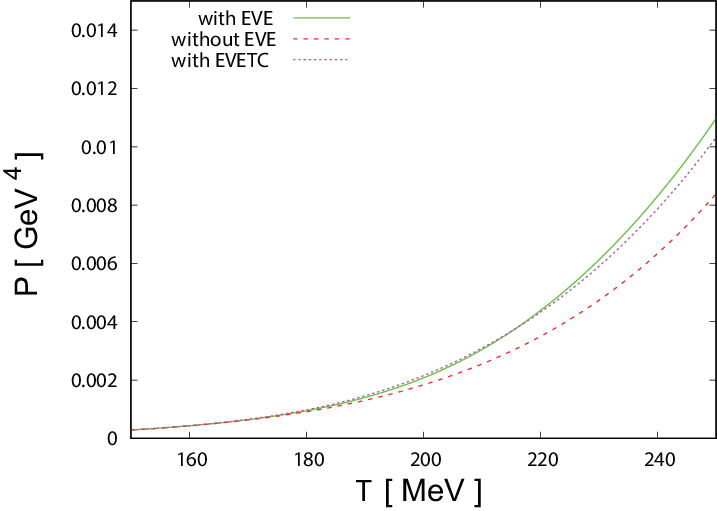}}
\caption{The $T$-dependence of the pressure when $\theta ={\pi\over{2}}$ and $b=v=4\pi r_0^3/3$. 
The solid, dashed, and dotted lines show the results obtained by the HRG models with EVE ($P$), without EVE ($P_{\rm p}$), with EVETC, respectively. 
}
 \label{Fig_P_theta=pi_2_comp}
\end{figure}

\begin{figure}[h]
\centering
\centerline{\includegraphics[width=0.40\textwidth]{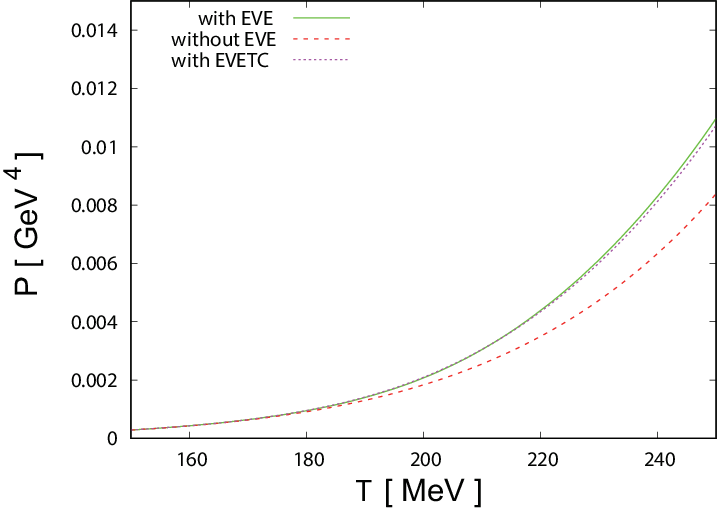}}
\caption{The $T$-dependence of the pressure when $\theta ={\pi\over{2}}$, $v=4\pi r_0^3/3$ and $b=1.315\,\mathrm{fm}^3$. 
The solid, dashed, and dotted lines show the results obtained by the HRG models with EVE ($P$), without EVE ($P_{\rm p}$), with EVETC, respectively. 
}
 \label{Fig_P_theta=pi_2_comp_2}
\end{figure}

In this subsection, we compare two formulations. 
Figure~\ref{Fig_P_mu=0_comp} shows the $T$-dependence of the pressure of the HRG model when $\mu=0$ and $b=v={4\pi r_0^3\over{3}}=2.14\,\mathrm{fm}^3$. 
We see that the exclude volume effects of EVE model is somewhat weaker than those of EVETC model, but the difference between two models is small. 
Figure~\ref{Fig_P_mu=0_comp_2} is similar to Fig.~\ref{Fig_P_mu=0_comp} but $b=1.315\,\mathrm{fm}^3$ is used.  
We see that the result of EVE almost coincides with that of EVETC. 
In this case, it seems that the deference of the two models can be eliminated by re-tuning the parameter $b$. 
\
\
Figure~\ref{Fig_P_mu=500_comp} shows $T$-dependence of the pressure of HRG model when $\mu=500$~MeV and $b=v={4\pi r_0^3\over{3}}$.  
We see that the difference between EVE and EVETC is small. 
Figure~\ref{Fig_P_mu=500_comp_2} is similar to Fig.~\ref{Fig_P_mu=500_comp} but $b=1.315\,\mathrm{fm}^3$ is used.  
Again, we see that the result of EVE almost coincides with that of EVETC.  
\
\
Next, we examine the results at imaginary $\mu =i\theta T$. 
Figure~\ref{Fig_P_theta=pi_2_comp} shows the $T$-dependence of the pressure of the HRG model when $\theta ={\pi\over{2}}$ and $b=v={4\pi r_0^3\over{3}}$.  
We see that the difference between EVE and EVETC is small. 
Fig.~\ref{Fig_P_theta=pi_2_comp_2} is similar to Fig.~\ref{Fig_P_theta=pi_2_comp} but $b=1.315\,\mathrm{fm}^3$ is used.  
The result of EVE almost coincides with that of EVETC. 
Comparison at the RW transition point will be shown in \S ~\ref{TPdep}. 
\
\

\subsection{Below $T_\mathrm{RW}$}

\begin{figure}[t]
\centering
\centerline{\includegraphics[width=0.40\textwidth]{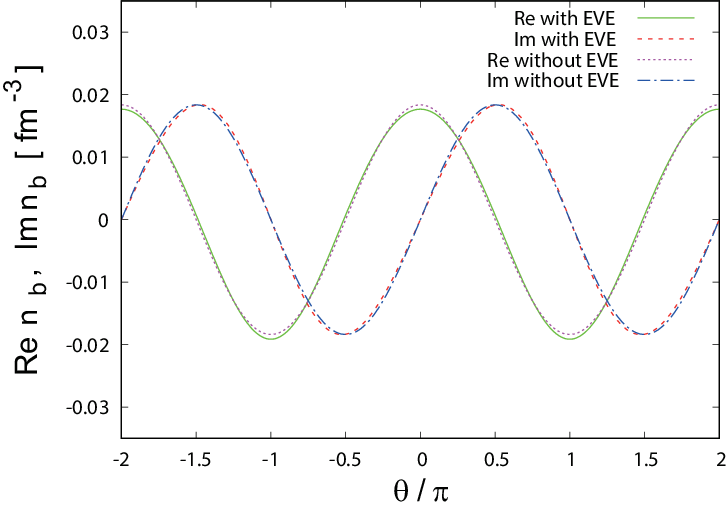}}
\caption{The $\theta$-dependence of the real and imaginary parts of number density of baryons when $T=150$~MeV. 
The solid, dashed, dotted and dot-dashed lines show the results of ${\rm Re}\,n_{\rm b}$, ${\rm Im}\,n_{\rm b}$, ${\rm Re}\,n_{\rm bp}$ and ${\rm Im}\,n_{\rm bp}$, respectively.  
 }
 \label{Fig_T=150_n_np}
\end{figure}

\begin{figure}[t]
\centering
\centerline{\includegraphics[width=0.40\textwidth]{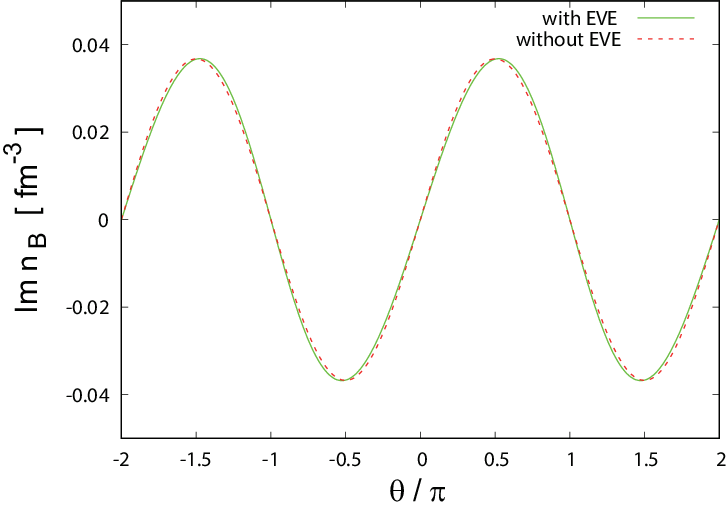}}
\caption{The $\theta$-dependence of imaginary part of baryon number density when $T=150$~MeV. 
The solid and dashed lines show the results of $n_{\rm B}$ and $n_{\rm Bp}$, respectively.  
 }
 \label{Fig_T=150}
\end{figure}

In this subsection, we show the $\theta$-dependence of the baryon number density and the pressure, and discuss the effects of EVE below $T_\mathrm{RW}$. 
Figure~\ref{Fig_T=150_n_np} shows the real and imaginary parts of $n_{\rm b}$ and $n_{\rm bp}$ as a function of $\theta$ when $T=150$~MeV. 
We see that the real (imaginary) part is an even (odd) function of $\theta$. 
We do not show the results of the anti-baryon contributions explicitly, since conditions (\ref{Ren}) and (\ref{Imn}) are satisfied.
As is expected, the relation ${\rm Re}[n_{\rm bp}(\theta)]={\rm Im}[n_{\rm bp}(\theta +{\pi\over{2}})]$ is approximately satisfied. 
The same relation is also approximately satisfied for $n_{\rm b}$ since EVE is negligible. 
Although we do not show all results explicitly, the relation (\ref{np_Re_Im}) is almost satisfied for $n_{\rm bp}$ and $n_{\rm ap}$ but is largely broken for $n_{\rm b}$ and $n_{\rm a}$, when the temperature is large and EVE is not negligible. 
We also remark that, as in the case with $T=150$~MeV, ${\rm Im}\,n_{\rm bp}={\rm Im}\,n_{\rm ap}=0$ and ${\rm Re}\,n_{\rm bp}={\rm Re}\,n_{\rm ap}<0$
are always satisfied at any temperature when $\theta =\pm\pi$. 
This is a very important property, as shown later.  
Figure~\ref{Fig_T=150} shows the imaginary part ${\rm Im}~n_{\rm B}$ of baryon number density as a function of $\theta$ when $T=150$~MeV. 
We see that EVE is negligible in this case. 

\begin{figure}[t]
\centering
\centerline{\includegraphics[width=0.40\textwidth]{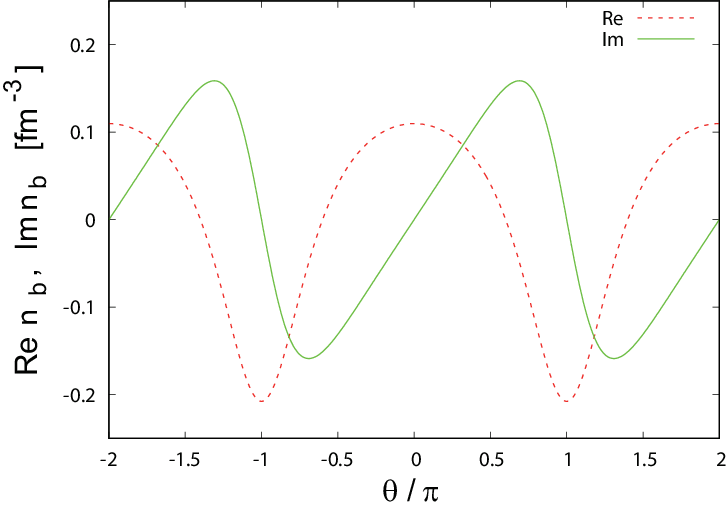}}
\caption{The $\theta$-dependence of the real and the imaginary part of number density $n_{\rm b}$ of baryons when $T=185$~MeV. 
The dashed and solid lines show the results for the real and the imaginary part, respectively.  
 }
\label{Fig_Renbe_Imnbe_T=185}
\end{figure}

\begin{figure}[h]
\centering
\centerline{\includegraphics[width=0.40\textwidth]{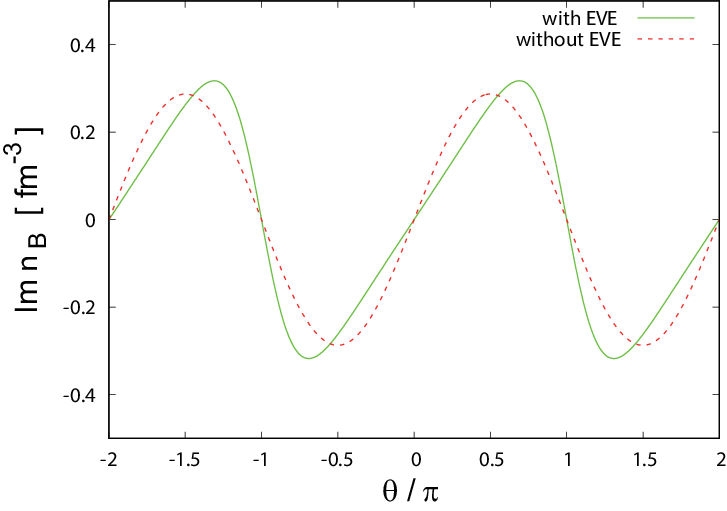}}
\caption{The $\theta$-dependence of the imaginary part of baryon number density when $T=185$~MeV. 
The solid and dashed lines show the results of $n_{\rm B}$ and $n_{\rm Bp}$, respectively.  
}
\label{Fig_T=185}
\end{figure}

\begin{figure}[h]
\centering
\centerline{\includegraphics[width=0.40\textwidth]{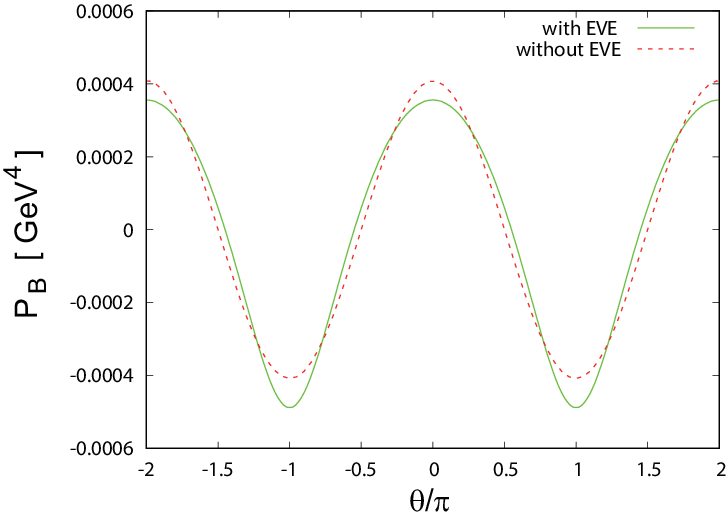}}
\caption{The $\theta$-dependence of the baryon pressure when $T=185$~MeV. 
The solid and dashed lines show the results with EVE ($P_{\rm B}$) and without EVE ($P_{\rm Bp}$), respectively. 
}
 \label{Fig_PB_T=185}
\end{figure}

Figure~\ref{Fig_Renbe_Imnbe_T=185} shows the real and the imaginary parts of $n_{\rm b}$ as a function of $\theta$ when $T=185$~MeV.  
We see that the equality ${\rm Re}[n_{\rm b}(\theta)]={\rm Im}[n_{\rm b}(\theta +{\pi\over{2}})]$ is broken largely by EVE. 
Figure~\ref{Fig_T=185} is the same as Fig.~\ref{Fig_T=150} but for $T=185$~MeV. 
Note that the temperature $T=185$~MeV is just below $T_{\rm RW}$ obtained by LQCD simulations. 
In this case, EVE is visible. 
These results indicate that the effects of EVE can be seen strongly at moderate $T$. 
Fig.~\ref{Fig_PB_T=185} shows the baryon pressure $P_{\rm B}$ as a function of $\theta$; we do not show the meson pressure since it does not depend on $\theta$ in our model. 
Both $P_{\rm B}$ and $P_{\rm Bp}$ are smooth functions of $\theta$. 
EVE is also visible in pressure.  

\subsection{Above $T_\mathrm{RW}$}

\begin{figure}[h]
\centering
\centerline{\includegraphics[width=0.40\textwidth]{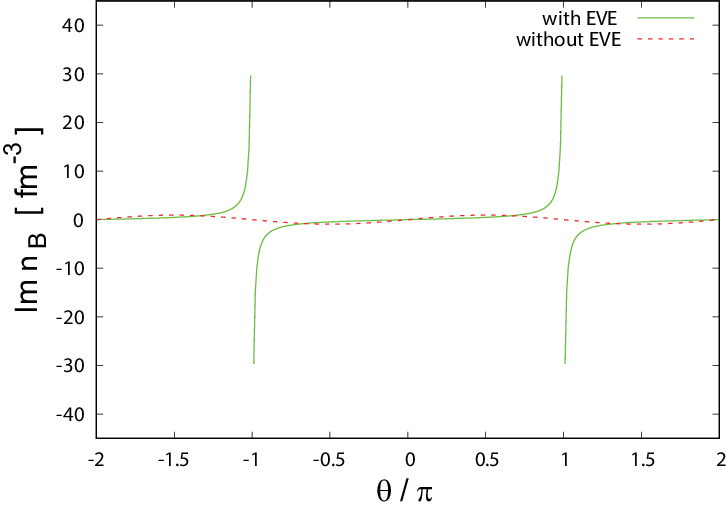}}
\caption{The $\theta$-dependence of the imaginary part of baryon number density when $T=210.8$~MeV. 
The solid and dashed lines show the results of $n_{\rm B}$ and $n_{\rm Bp}$, respectively.  
 }
 \label{Fig_T=210_8} 
\end{figure}

\begin{figure}[h]
\centering
\centerline{\includegraphics[width=0.40\textwidth]{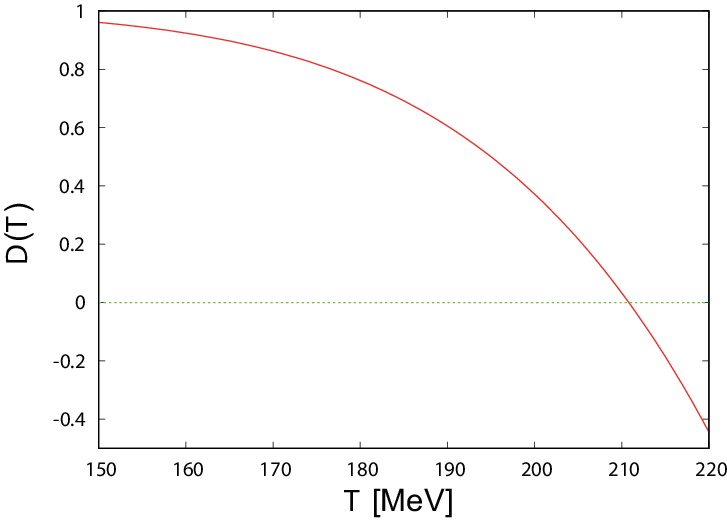}}
\caption{The $T$-dependence of $D(T)$ at $\theta =\pi$ (solid line). 
The dotted line shows the condition $D(T)=0$. 
 }
 \label{Fig_DT}
\end{figure}

Figure~\ref{Fig_T=210_8} is the same as Fig.~\ref{Fig_T=150} but for $T=210.8$~MeV. 
In this case, ${\rm Im}\,n_{\rm B}$ seems to be discontinuous at $\theta =\pm \pi$. 
The result resembles the one obtained by the PNJL model in the bottom panel of Fig.~\ref{Fig_T=150_250_PNJL}. 
However, different from the PNJL case, the height of the peak (the depth of the negative peak) increases as the minimal interval $\Delta \theta$ in the horizontal axis decreases in numerical calculation.   
It is very interesting that the temperature $T=210.8$~MeV is slightly larger than $T_{\rm RW}$ obtained by LQCD simulations. 
As is mentioned above, ${\rm Im}\,n_{\rm bp}={\rm Im}\,n_{\rm ap}=0$ and ${\rm Re}\,n_{\rm bp}={\rm Re}\,n_{\rm ap}<0$ are always satisfied at any temperature when $\theta =\pm\pi$. 
Hence $\theta =\pm \pi$, the first and third ones of Eq.\,(\ref{n_form}) reduce to 
\begin{eqnarray}
{\rm Re}\,n_{\rm b}
={{\rm Re}\,n_{\rm bp}\over{D(T)}}
={\rm Re}\,n_{\rm a}
={{\rm Re}\,n_{\rm ap}\over{D(T)}},
\label{n_form_2}
\end{eqnarray}
when  
\begin{eqnarray}
D(T)=1+v\,{\rm Re}\,n_{\rm bp}=1+v\,{\rm Re}\,n_{\rm ap},
\label{DT}
\end{eqnarray}
does not vanish. 
Figure~\ref{Fig_DT} shows $D(T)$ as a function of $T$ when $\theta =\pi$. 
$D(T)=0$ at $T= 210.8$~MeV. 
This means ${\rm Re}~n_{\rm bp}={\rm Re}~n_{\rm ap}=-1/v$.
The vanishing of $D(T)$ makes $n_{\rm B}$ singular at $T=210.8$~MeV. 

If we define $n^\prime =-{\rm Re}~n_{\rm b}=-{\rm Re}~n_{\rm a}$ and $n^\prime_{\rm p}=-{\rm Re}~n_{\rm bp}=-{\rm Re}~n_{\rm ap}>0$, Eq.~(\ref{n_form_2}) can be rewritten as
\begin{eqnarray}
n^\prime ={n^\prime_{\rm p}\over{1-v\,n^\prime_{\rm p}}}. 
\label{n_form_3_1}
\end{eqnarray}
Hence, we obtain
\begin{eqnarray}
n^\prime_{\rm p} ={n^\prime\over{1+v\,n^\prime}}. 
\label{n_form_3}
\end{eqnarray}
This is nothing but the dual EVE (\ref{EVE_1_dual}). 
If $n^\prime\to \infty$, $n^\prime_{\rm p} \to 1/v$.  
Figure~\ref{Fig_T_nprime} shows $n^\prime$ as a function of $T$ when $\theta =\pi$. 
$n^\prime$ has a divergent behavior when $T\to 210.8$~MeV where $n^\prime_{\rm p}=1/v$; note that the height of the peak (the depth of the negative peak) increases as the minimal interval $\Delta T$ in horizontal axis decreases in numerical calculation. 
In this paper, we call this situation "dual dense packing limit" of baryons. 
The dual dense packing of baryons causes the singularity; note that, although ${\rm Re}\,n_{\rm B}$ always vanishes, the effects of ${\rm Re}\,n_{\rm bp}$ and ${\rm Re}\,n_{\rm ap}$ remain in other thermodynamic quantities such as ${\rm Im}\,n_{\rm B}$ and pressure.   

We also note that, when $n^\prime <0$, the interpretation of the dual EVE is not appropriate in ({\ref{n_form_3}). 
However, $n^\prime_{\rm p} \to 1/v$ is satisfied, when $n^\prime \to -\infty$. 
Similarly, since ${\rm Re}~n_{\rm bp}<0$, the interpretation of the EVE is not appropriate in ({\ref{n_form_2}), but ${\rm Re}\,n_{\rm b} \to 1/v$ is satisfied when ${\rm Re}~n_{\rm bp} \to -\infty$. 
Hence, $1/v$ is the high temperature limit of ${\rm Re}\,n_{\rm b}$ at $\theta =\pm\pi$.  

\begin{figure}[t]
\centering
\centerline{\includegraphics[width=0.40\textwidth]{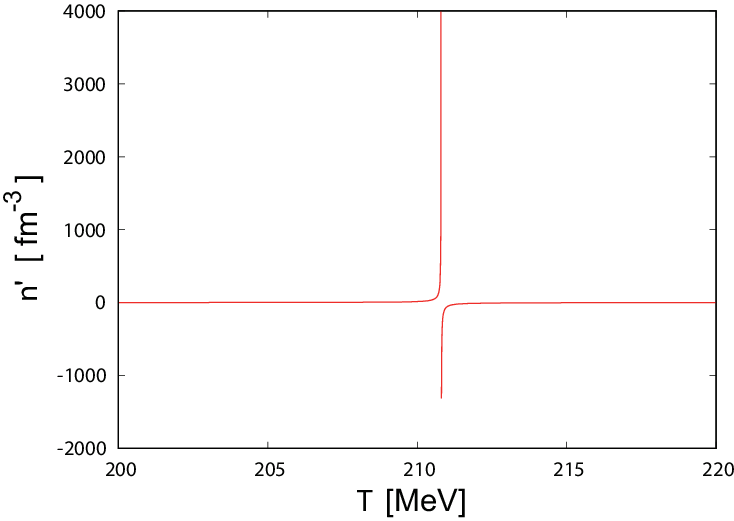}}
\caption{The $T$-dependence of $n^\prime$ at $\theta =\pi$. 
}
\label{Fig_T_nprime}
\end{figure}

When the Boltzmann distribution approximation is valid, 
we can define a $\mu$-independent effective baryon radius as follows. 
In this approximation, $D(T)$ is reduced to $1-v/v^*(T)$ where 
\begin{eqnarray}
v^*(T)={1\over{A(T)}},
\label{VT}
\end{eqnarray}
with
\begin{eqnarray}
A(T)=\sum_{i}A_i. 
\label{Asum}
\end{eqnarray}
Note that $A(T)$ is the amplitude of $n_{\rm bp}$ and $n_{\rm bp}$ when they oscillate as $\theta$ varies.     
Figure~\ref{Fig_rT} shows an effective baryon radius $r^*(T)= [3v^*/(4\pi) ]^{1/3}$ as a function of $T$. 
We see that $r^*(T)\sim r_0=0.8$~fm at $T\sim 211$~MeV. 
This means that the Boltzmann distribution approximation is valid in this case. 

\begin{figure}[t]
\centering
\centerline{\includegraphics[width=0.40\textwidth]{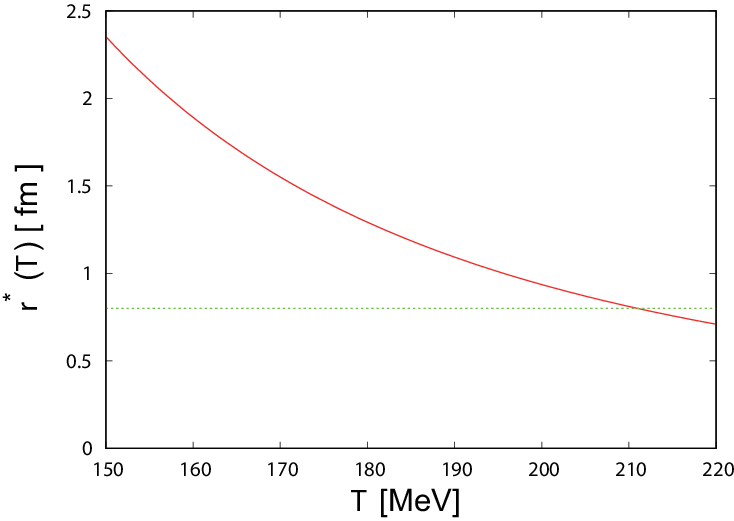}}
\caption{The $T$-dependence of the effective baryon radius  $r^*(T)$ at $\theta =\pi$ (solid line).
The dotted line shows the condition $r^*=r_0=0.8$~fm. 
 }
 \label{Fig_rT}
\end{figure}

Figure~\ref{Fig_theta_PB_T=210_8} is the same as Fig.~\ref{Fig_PB_T=185} but for $T=210.8$~MeV. 
$P_{\rm B}$ has a cusp at $\theta =\pm \pi$, while $P_{\rm Bp}$ does not. The result of $P_{\rm B}$ resembles the one obtained by the PNJL model in the bottom panel of Fig.~\ref{Fig_PNJL_P}. 
However, different from the PNJL case, 
there is a tendency that the depth of the negative peak increases as the minimal interval $\Delta \theta$ in the horizontal axis decreases in numerical calculation. 
The cusp in $P_{\rm B}$ causes the discontinuity in ${\rm Im}~n_{\rm B}$. 
\begin{figure}[h]
\centering
\centerline{\includegraphics[width=0.40\textwidth]{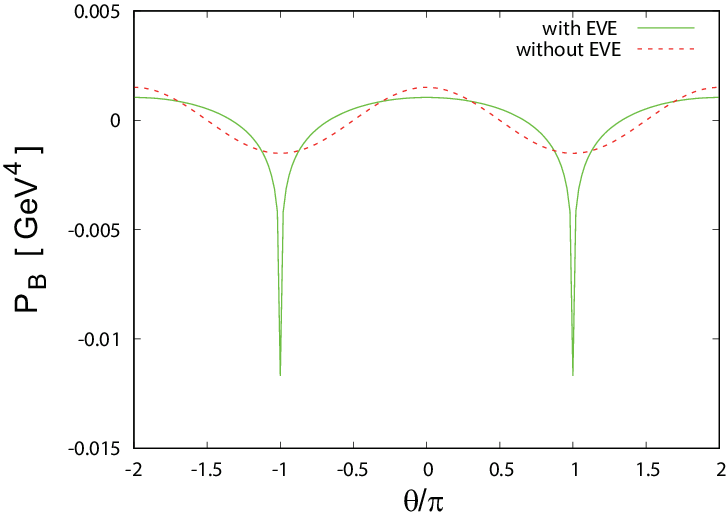}}
\caption{The $\theta$-dependence of the baryon pressure when $T=210.8$~MeV. 
The solid and dashed lines show the results with EVE ($P_{\rm B}$) and without EVE ($P_{\rm Bp}$), respectively. 
}
 \label{Fig_theta_PB_T=210_8}
\end{figure}
Figure~\ref{Fig_T=250} is the same as Fig.~\ref{Fig_T=150} but for $T=250$~MeV. 
In this case, ${\rm Im}~n_{\rm b}$ is a smooth function of $\theta$ as in the cases at low temperature. 
This feature is different from the PNJL result in Fig.~\ref{Fig_T=150_250_PNJL} where the discontinuities appear when $T>T_{\rm RW}$. 
\begin{figure}[t]
\centering
\centerline{\includegraphics[width=0.40\textwidth]{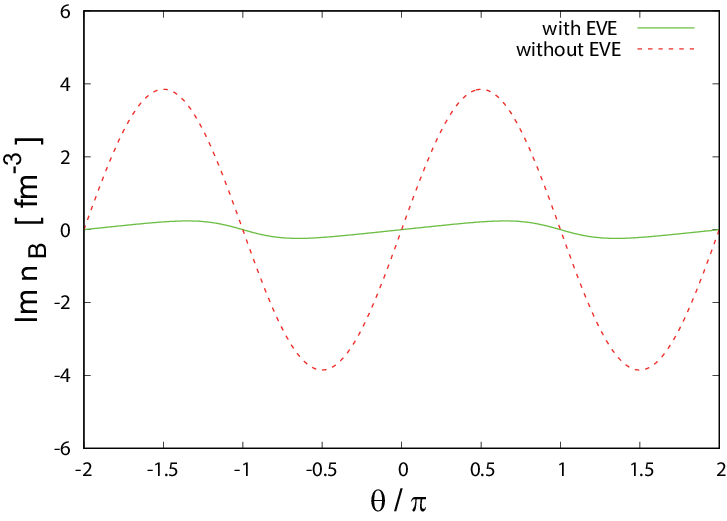}}
\caption{The $\theta$-dependence of the imaginary part of baryon number density when $T=250$~MeV. 
The solid and dashed lines show the results of $n_{\rm B}$ and $n_{\rm Bp}$, respectively.  
 }
 \label{Fig_T=250}
\end{figure}
Figure~\ref{Fig_PB_T=250} is the same as Fig.~\ref{Fig_PB_T=185} but for $T=250$~MeV. 
Again, both $P_{\rm B}$ and $P_{\rm Bp}$ are smooth functions of $\theta$. 
As is in the case with ${\rm Im}\,n_{\rm B}$, this feature is different from the PNJL result in Fig.~\ref{Fig_PNJL_P} where the cusps appear when $T>T_{\rm RW}$. 
The difference seems to be originated in the lack of quark degree of freedom in HRG model.  
\begin{figure}[t]
\centering
\centerline{\includegraphics[width=0.40\textwidth]{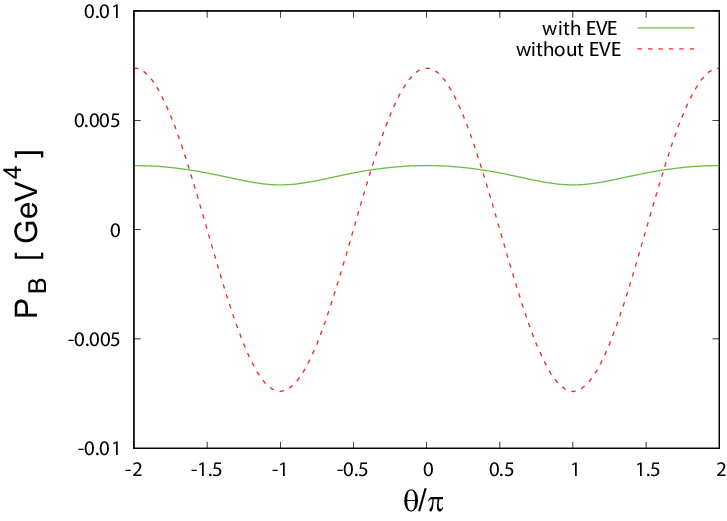}}
\caption{The $\theta$-dependence of the baryon pressure when $T=250$~MeV. 
The solid and dashed lines show the results with EVE ($P_{\rm B}$) and without EVE ($P_{\rm Bp}$), respectively. 
 }
 \label{Fig_PB_T=250}
\end{figure}

\subsection{$T$-dependence of pressure}
\label{TPdep}

Figure~\ref{Fig_P_theta=pi} shows the hadron pressure as a function of $T$ at $\theta =\pi$. 
Note that the meson contribution is included in the hadron pressure. 
The pressure $P$ in the HRG model with EVE has a cusp at $T\sim 210.8$~MeV, while $P_{\rm p}$ in the HRG model without EVE is a smooth function of $T$; 
there is a tendency that the depth of the negative peak of $P$ ($P^\prime$) increases as the minimal interval $\Delta T$ in the horizontal axis decreases in the numerical calculation.  
$P_{\rm p}$ increase as temperature increase when $T<230$~MeV, but decreases $T>230$~MeV. 
This is because the baryon contribution is negative at $\theta =\pi$ and the decrease of the negative baryon contribution overcomes the increase of the positive meson contribution when $T>230$~MeV. 
On the contrary, $P$ increases as $T$ increases in the high temperature region, although it has a negative value in the intermediate temperature region. 
\begin{figure}[h]
\centering
\centerline{\includegraphics[width=0.40\textwidth]{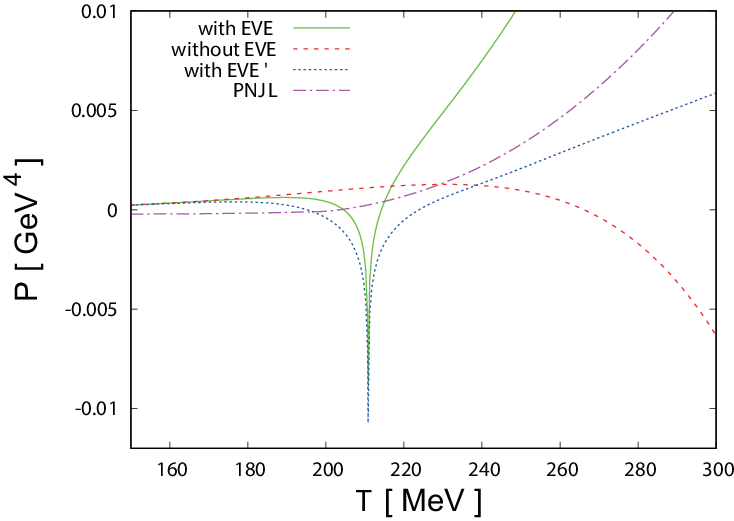}}
\caption{The $T$-dependence of the pressure when $\theta =\pi$. 
The solid, dashed, dotted and dot-dashed lines show the results obtained by the HRG models with EVE ($P$), without EVE ($P_{\rm p}$), with EVE plus meson gas suppression ($P^\prime =P_{\rm B}+P_{\rm M}^\prime $), and the PNJL model, respectively. 
}
 \label{Fig_P_theta=pi}
\end{figure}\

In Fig.~\ref{Fig_P_theta=pi}, the result obtained by the PNJL model is also shown. 
We have subtracted $P_{\rm PNJL}(T=0,\mu =0)+B$ from $P_{\rm }(T,\mu)$ where $B=(150~{\rm MeV})^4$ is the bag constant.  
$P_{\rm PNJL}$ is larger than $P_{\rm p}$ when $T>229$~MeV, while it is smaller than $P_{\rm p}$ when $T<229$~MeV.  
If we apply the Gibbs criteria to this case without EVE, the quark phase (hadron phase) is realized when $T>229$~MeV ($T<229$~MeV). 

$P_{\rm PNJL}$ is larger than $P$ in the region $T=204\sim 214$~MeV, while it is smaller in the other regions. 
If we apply the Gibbs criteria to this case with EVE, the quark phase is realized in the region $T=204\sim 214$~MeV and the hadron phase is realized otherwise. 
However, the HRG model with EVE may be invalid when $T>210.8$~MeV, since it has a singularity. 
If this interpretation is appropriate, the quark phase is realized when $T>204$~MeV. 

It should be noted that the rapid increase of pressure of the HRG model with EVE in the high temperature region is caused by the rapid increase of meson contribution. 
In our model, the meson gas is treated as a free gas of point-like particles.  
It is natural that the meson contribution is also suppressed in the high temperature region. 
For illustration, we assume the following simple suppression form of the meson pressure.  
\begin{eqnarray}
P_{\rm M}^\prime =P_{\rm M}\exp{[-(T/T^*)^{10}]}
\label{PMprime}
\end{eqnarray}
We set $T^*=210.8$~MeV. 
As is seen in Fig.~\ref{Fig_P_theta=pi}, in this case, the hadron phase is realized in the low temperature and the quark phase is realized in the high temperature region ($T>197$~MeV). 
The study of the mechanism of meson gas suppression is an important subject in the future.

Figure~\ref{Fig_P_theta=pi_comp} shows $T$-dependence of the pressure of HRG model when $\theta =\pi$ and $b=v={4\pi r_0^3\over{3}}=2.14 \, \mathrm{fm}^3$. 
In the case of EVETC, we found the singular behavior at $T=$190.5~MeV, but could not find the solution when $T>$190.5~MeV. 
The temperature at which singularity occurs is somewhat smaller in EVETC than EVE. 
Figure~\ref{Fig_P_theta=pi_comp_2} is similar to Fig.~\ref{Fig_P_theta=pi_comp} but $b=1.315 \, \mathrm{fm}^3$ is used. 
The difference of the singularity temperature is somewhat smaller than that in Fig.~\ref{Fig_P_theta=pi_comp}. 
As is in Fig.~\ref{Fig_P_theta=pi_comp}, we could not find the solution when $T>$201~MeV. 
It seems that the singularity temperature and the property of the pressure above the singularity temperature depend on the detail description of the model to some degree. 
However, qualitatively, the result of EVE is consistent with that of EVETC below the singularity temperature. 
\begin{figure}[h]
\centering
\centerline{\includegraphics[width=0.40\textwidth]{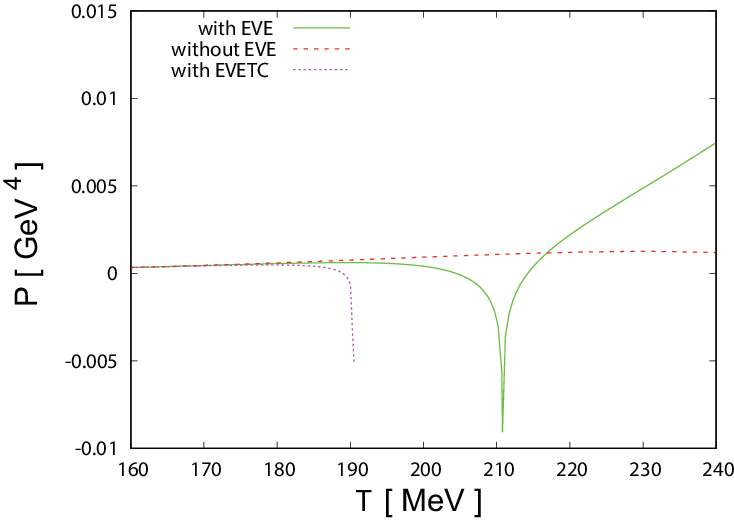}}
\caption{The $T$-dependence of the pressure when $\theta =\pi$ and $b=v=4\pi r_0^3/3$. 
The solid, dashed, and dotted lines show the results obtained by the HRG models with EVE ($P$), without EVE ($P_{\rm p}$), with EVETC, respectively. 
}
 \label{Fig_P_theta=pi_comp}
\end{figure}\
\begin{figure}[h]
\centering
\centerline{\includegraphics[width=0.40\textwidth]{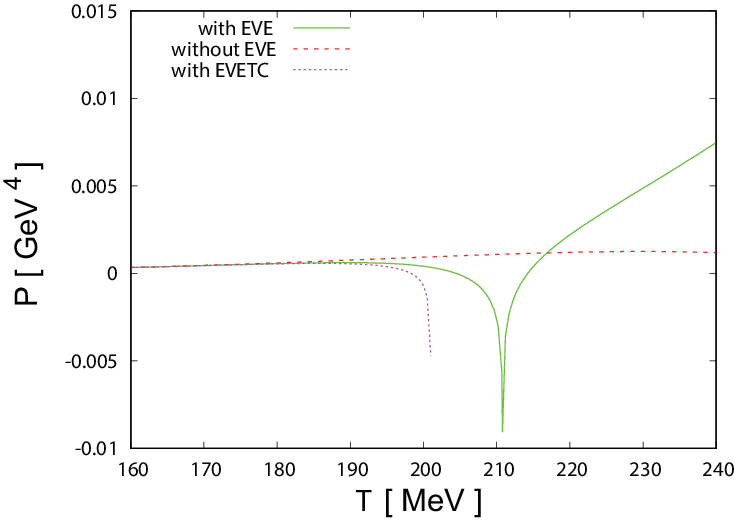}}
\caption{The $T$-dependence of the pressure when $\theta =\pi$, $v=4\pi r_0^3/3$ and $b=1.315\,\mathrm{fm}^3$. 
The solid, dashed, and dotted lines show the results obtained by the HRG models with EVE ($P$), without EVE ($P_{\rm p}$), with EVETC, respectively. 
}
 \label{Fig_P_theta=pi_comp_2}
\end{figure}\

Figure~\ref{Fig_P_theta=0} shows the same as Fig.~\ref{Fig_P_theta=pi} but for $\mu =0$. 
We see that $P_{\rm p}$ and $P$ are larger than $P_{\rm PNJL}$ at high temperature. 
Hence, if we apply the Gibbs criteria to these cases, the hadron phase is realized at high temperature. 
On the contrary, $P^\prime$ is smaller than $P_{\rm PNJL}$ at high temperature. 
In this case, the quark phase is realized at high temperature ($T>216$~MeV).  
This fact confirms the importance of the meson pressure suppression.  
\begin{figure}[t]
\centering
\centerline{\includegraphics[width=0.40\textwidth]{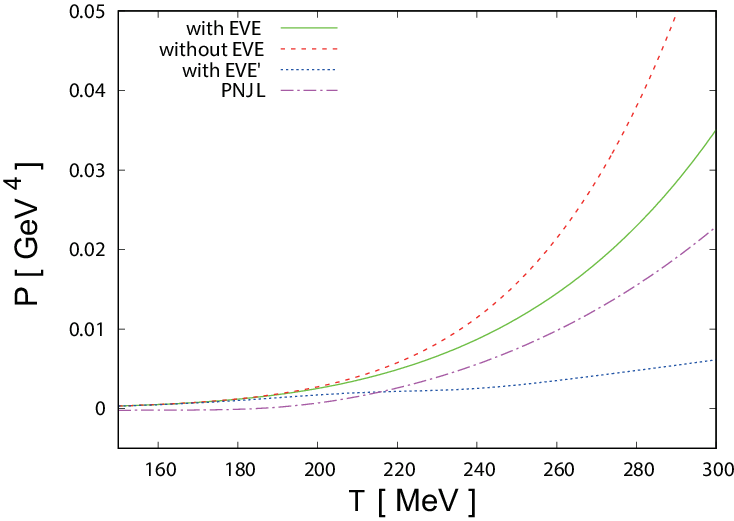}}
\caption{The $T$-dependence of the hadron (PNJL) pressure when $\mu =0$.
The solid, dashed, dotted and dot-dashed lines show the results obtained by the HRG models with EVE ($P$), without EVE ($P_{\rm p}$), with EVE plus meson gas suppression ($P^\prime =P_{\rm B}+P_{\rm M}^\prime $), and the PNJL model, respectively. 
}
 \label{Fig_P_theta=0}
\end{figure}

 \section{Summary}
\label{summary}

In summary, in this paper, we have studied the baryon number chemical potential ($\mu$) dependence of the baryon number density and pressure in the HRG model with excluded volume effects (EVE) when $\mu (=i\theta T)$ is pure imaginary. 
We compare the results with the ones obtained by the PNJL model in which the quark degree of freedom is contained. 

In the HRG model, the Roberge-Weiss (RW) periodicity, which the QCD grand canonical partition function has, is trivial.
At low temperature and high temperature, the $\theta$-dependence of the baryon number density and the pressure are smooth functions of $\theta$. 
However, they have a singular behavior at $\theta =(2k+1)\pi$ with $k \in \mathbb{Z}$ 
when $T\sim 211$~MeV which is consistent with the RW transition temperature $T_{\rm RW}$ obtained by lattice QCD (LQCD) simulations. 
This coincidence may reflect the simple fact that the inverse of the baryon radius ${1 / r_0}$ is roughly estimated as $200$~MeV. 
As $T$ increases, the equality ${\rm Re}(n_{\rm b}(\theta ))
={\rm Im}(n_{\rm a}(\theta +{\pi\over{2}}))$ is largely broken by EVE. 
This breaking of the equality seems to be important for the singular behavior at $T\sim 211$~MeV. 
It is very interesting that the singularity is well explained by the dual EVE in which the roles of point-like and non point-like particles are exchanged with each other in EVE.
In the picture of the dual EVE, the dual dense packing causes the singularity. 
At high temperature above $T\sim 211$~MeV, the singularity disappears. 
However, it is natural that the HRG model with EVE is valid only when the temperature is smaller than the temperature where the singularity appears. 
Since the dual EVE is equivalent to the EVE with a negative particle volume, it may be regarded as attractive force effects. 
It considerably lowers the pressure. 
Hadron matter with pressure of negative infinity cannot exist.  
Hadron matter is compressed and baryons are crushed by the strong attractive force at $\theta =(2k+1)\pi$ when $T\sim 211$~MeV. 
Hence, the introduction of the quark degree of freedom is necessary at higher temperature. 
The thermodynamic comparison between the HRG results and the PNJL results ensures this expectation. 

It is also very interesting that EVE is visible just before $T_{\rm RW}$. 
EVE may be detectable by the LQCD simulation at imaginary $\mu$~\cite{Vovchenko:2017xad}. 
If the function form of ${\rm Im}\,n_{\rm B}$ deviates from the sine function, the EVE may be significant. 
In our calculations, we have treated $v$ as a constant parameter. 
However, the $T$ and/or the $\mu$-dependence may be important; see, e.g., Ref.~\cite{Kouno:2023ygw}.  
Such a dependence may be determined by the LQCD simulations. 


\begin{acknowledgments}
H. K. thanks A. Miyahara, M. Ishii and M. Yahiro for useful discussions on the excluded volume effects and the RW periodicity. 
\end{acknowledgments}

\appendix

\section{PNJL model}
\label{PNJLmodel}

Thermodynamic potential density $\Omega_{\rm PNJL}$ in the three flavor PNJL model with the mean field approximation is given by ~\cite{Fukushima:2003fw,Ratti:2005jh,Ghosh:2006qh,Megias:2004hj,Roessner:2006xn,Sakai:2010rp}
\begin{eqnarray}
&&\Omega_{\rm PNJL}(T,\mu_{\rm q})={\cal U}(T)+U_{\rm M}+\sum_{f=u,d,s}\Omega_{f}(T,\mu_{\rm q}),
\nonumber\\
\label{Omega_PNJL}
\end{eqnarray} 
where
\begin{align}
\Omega_f(T,\mu_{\rm q}) &=
-2\int {d^3p\over{(2\pi )^3}}~\Big[3\sqrt{p^2+M_f^2}
\nonumber\\
& +T \log{ \Bigl[ f^-_f(p) \, f^+_f(p) \Bigr] }\Big], 
\label{Omega_PNJL_f_g}
\end{align} 
with
\begin{align}
f^{-}_f (p)
&= 1 + 3 \Bigl[ \Phi + \Phi^*e^{-\beta E^{-}_f} \Bigl] e^{-\beta E^{-}_f} + e^{-3\beta E^{-}_f}, 
\nonumber\\
f^{+}_f (p)
&= 1 + 3 \Bigl[ \Phi^* + \Phi e^{-\beta E^{+}_f} \Bigr] e^{-\beta E^{+}_f} + e^{-3\beta E^{+}_f}. 
\label{Omega_PNJL_f}
\end{align} 
where $\Phi$, ${\cal U}$, $U_{\rm M}$ and $M_f$ are the Polyakov loop, the Polyakov-loop potential, the mesonic potential and the effective mass of $f$-quark, respectively, and $E^\pm_f(p)=\sqrt{p^2+M_f^2}\pm \mu_{\rm q}$. 
The first term of the integrand in Eq.\,(\ref{Omega_PNJL_f_g}) is a vacuum contribution. 
Note that the Polyakov-loop $\Phi$ and its conjugate $\Phi^*$ are complex, in general.   
The effective quark mass $M_f$ is given by 
\begin{eqnarray}
M_{f}&=&m_{f}-4G_{\rm S}\sigma_f+2G_{\rm D}\sigma_f^\prime\sigma_f^{\prime\prime},
\label{quakrmass}
\end{eqnarray}
with
\begin{align}
f\neq f^\prime,~f\neq f^{\prime\prime},~f^\prime \neq f^{\prime\prime}, 
\label{flavor}
\end{align}
where $m_{f}$ is the current quark mass of $f$-quark, $G_{\rm S}$ and $G_{\rm D}$ are coupling constants of four and six quarks interaction, and $\sigma_f$ is the chiral condensate of $f$-quark, respectively. 
The mesonic part is given by
\begin{eqnarray}
U_{\rm M}=2G_{\rm S}(\sigma_u^2+\sigma_d^2+\sigma_s^2)-4G_{\rm D}\sigma_u\sigma_d\sigma_s. 
\label{U_mesonic}
\end{eqnarray}
According to Ref.~\cite{Rehberg:1995kh}, for model parameters, we set $m_{u,d}=5.5$~MeV, $m_s=140.7$~MeV, $G_{\rm S}\Lambda^2=1.835$, $G_{\rm D} \Lambda^5=12.36$ and $\Lambda =602.3$~MeV. 
It should be noted that, to ensure the conformal limit at high temperature, we introduce a momentum cutoff $\Lambda$ only in the vacuum part; the original PNJL model has a cutoff because it is a non-renormalizable model. 
For the Polyakov potential, we use the following form~\cite{Roessner:2006xn};
\begin{eqnarray}
&&\frac{{\cal U}(T; \Phi,\Phi^*)}{T^4}
= \Big[-{a(T)\over{2}}\Phi^*\Phi
\nonumber\\
&&+b(T)\log{ \{ 1-6\Phi^*\Phi+4(\Phi^3+{\Phi^*}^3)-3(\Phi^*\Phi)^2 \} } \Big]
\nonumber\\
\label{PP}
\end{eqnarray}
where
\begin{eqnarray}
a(T)&=&a_0+a_1\left({T_0\over{T}}\right)+a_2\left({T_0\over{T}}\right)^2,
\label{PPpara_a}
\\
b(T)&=&b_3\left({T_0\over{T}}\right)^3,
\label{PPpara_b}
\end{eqnarray}
with modified coefficients~\cite{Miyahara:2017eam,Miyahara:2019zfn}
\begin{eqnarray}
a_0&=&2.457,~~a_1=-2.47,~~a_2=15.2,~~b_3=-1.75. 
\nonumber\\
\label{PPpara}
\end{eqnarray}
In numerical calculations, we set $T_0=220$~MeV so as to satisfy $T_{\rm RW}\sim 200$~MeV.  

The pressure in the PNJL model is given by $P_{\rm PNJL}=-\Omega_{\rm PNJL}$. 
The number densities of 
$f$-quark and anti-quark ($n_{{\rm q},f}$ and $n_{{\rm aq},f}$) are given by
\begin{align}
n_{{\rm q},f} &= {1\over{\pi^2}}\int_0^\infty dp~p^2~{g^{-}_f(p)\over{f^{-}_f(p)}},
\label{nqf}
\\
n_{{\rm aq},f} &= {1\over{\pi^2}}\int_0^\infty dp~p^2~{g^{+}_f(p)\over{f^{+}_f(p)}}, 
\label{naqf}
\end{align}
where  
\begin{eqnarray}
g^{-}_f (p)=3\Phi e^{-\beta E^{-}_f}+6\Phi^*e^{-2\beta E^{-}_f}+3e^{-3\beta E^{-}_f}, 
\nonumber\\
\label{gm}
\\
g^{+}_f (p)=3\Phi^* e^{-\beta E^{+}_f}+6\Phi e^{-2\beta E^{+}_f}+3e^{-3\beta E^{+}_f}. 
\nonumber\\
\label{gp}
\end{eqnarray}
The solutions of $\sigma_f~(f=u,d,s)$, $\Phi$ and $\Phi^*$ are determined so as to minimize $\Omega_{\rm PNJL}$.

\bibliography{ref.bib}

\begin{thebibliography}{35}%
\makeatletter
\providecommand \@ifxundefined [1]{%
 \@ifx{#1\undefined}
}%
\providecommand \@ifnum [1]{%
 \ifnum #1\expandafter \@firstoftwo
 \else \expandafter \@secondoftwo
 \fi
}%
\providecommand \@ifx [1]{%
 \ifx #1\expandafter \@firstoftwo
 \else \expandafter \@secondoftwo
 \fi
}%
\providecommand \natexlab [1]{#1}%
\providecommand \enquote  [1]{``#1''}%
\providecommand \bibnamefont  [1]{#1}%
\providecommand \bibfnamefont [1]{#1}%
\providecommand \citenamefont [1]{#1}%
\providecommand \href@noop [0]{\@secondoftwo}%
\providecommand \href [0]{\begingroup \@sanitize@url \@href}%
\providecommand \@href[1]{\@@startlink{#1}\@@href}%
\providecommand \@@href[1]{\endgroup#1\@@endlink}%
\providecommand \@sanitize@url [0]{\catcode `\\12\catcode `\$12\catcode `\&12\catcode `\#12\catcode `\^12\catcode `\_12\catcode `\%12\relax}%
\providecommand \@@startlink[1]{}%
\providecommand \@@endlink[0]{}%
\providecommand \url  [0]{\begingroup\@sanitize@url \@url }%
\providecommand \@url [1]{\endgroup\@href {#1}{\urlprefix }}%
\providecommand \urlprefix  [0]{URL }%
\providecommand \Eprint [0]{\href }%
\providecommand \doibase [0]{http://dx.doi.org/}%
\providecommand \selectlanguage [0]{\@gobble}%
\providecommand \bibinfo  [0]{\@secondoftwo}%
\providecommand \bibfield  [0]{\@secondoftwo}%
\providecommand \translation [1]{[#1]}%
\providecommand \BibitemOpen [0]{}%
\providecommand \bibitemStop [0]{}%
\providecommand \bibitemNoStop [0]{.\EOS\space}%
\providecommand \EOS [0]{\spacefactor3000\relax}%
\providecommand \BibitemShut  [1]{\csname bibitem#1\endcsname}%
\let\auto@bib@innerbib\@empty
\bibitem [{\citenamefont {Fukushima}\ and\ \citenamefont {Hatsuda}(2011)}]{Fukushima:2010bq}%
  \BibitemOpen
  \bibfield  {author} {\bibinfo {author} {\bibfnamefont {K.}~\bibnamefont {Fukushima}}\ and\ \bibinfo {author} {\bibfnamefont {T.}~\bibnamefont {Hatsuda}},\ }\href {\doibase 10.1088/0034-4885/74/1/014001} {\bibfield  {journal} {\bibinfo  {journal} {Rept. Prog. Phys.}\ }\textbf {\bibinfo {volume} {74}},\ \bibinfo {pages} {014001} (\bibinfo {year} {2011})},\ \Eprint {http://arxiv.org/abs/1005.4814} {arXiv:1005.4814 [hep-ph]} \BibitemShut {NoStop}%
\bibitem [{\citenamefont {de~Forcrand}(2009)}]{deForcrand:2010ys}%
  \BibitemOpen
  \bibfield  {author} {\bibinfo {author} {\bibfnamefont {P.}~\bibnamefont {de~Forcrand}},\ }\href@noop {} {\bibfield  {journal} {\bibinfo  {journal} {PoS}\ }\textbf {\bibinfo {volume} {LAT2009}},\ \bibinfo {pages} {010} (\bibinfo {year} {2009})},\ \Eprint {http://arxiv.org/abs/1005.0539} {arXiv:1005.0539 [hep-lat]} \BibitemShut {NoStop}%
\bibitem [{\citenamefont {Nagata}(2020)}]{Nagata:2021bru}%
  \BibitemOpen
  \bibfield  {author} {\bibinfo {author} {\bibfnamefont {K.}~\bibnamefont {Nagata}},\ }\href {https://www2.yukawa.kyoto-u.ac.jp/~soken.editorial/sokendenshi/vol31/Nagata2016GKH_EI_revised.pdf} {\bibfield  {journal} {\bibinfo  {journal} {\begin{CJK}{UTF8}{ipxm}素粒子論研究\end{CJK} (Soryusironkenkyu)}\ }\textbf {\bibinfo {volume} {31}},\ \bibinfo {pages} {1} (\bibinfo {year} {2020})}\BibitemShut {NoStop}%
\bibitem [{\citenamefont {Nagata}(2022)}]{Nagata:2021ugx}%
  \BibitemOpen
  \bibfield  {author} {\bibinfo {author} {\bibfnamefont {K.}~\bibnamefont {Nagata}},\ }\href {\doibase 10.1016/j.ppnp.2022.103991} {\bibfield  {journal} {\bibinfo  {journal} {Prog. Part. Nucl. Phys.}\ }\textbf {\bibinfo {volume} {127}},\ \bibinfo {pages} {103991} (\bibinfo {year} {2022})},\ \Eprint {http://arxiv.org/abs/2108.12423} {arXiv:2108.12423 [hep-lat]} \BibitemShut {NoStop}%
\bibitem [{\citenamefont {de~Forcrand}\ and\ \citenamefont {Philipsen}(2002)}]{deForcrand:2002ci}%
  \BibitemOpen
  \bibfield  {author} {\bibinfo {author} {\bibfnamefont {P.}~\bibnamefont {de~Forcrand}}\ and\ \bibinfo {author} {\bibfnamefont {O.}~\bibnamefont {Philipsen}},\ }\href {\doibase 10.1016/S0550-3213(02)00626-0} {\bibfield  {journal} {\bibinfo  {journal} {Nucl.Phys.}\ }\textbf {\bibinfo {volume} {B642}},\ \bibinfo {pages} {290} (\bibinfo {year} {2002})},\ \Eprint {http://arxiv.org/abs/hep-lat/0205016} {arXiv:hep-lat/0205016 [hep-lat]} \BibitemShut {NoStop}%
\bibitem [{\citenamefont {D'Elia}\ and\ \citenamefont {Lombardo}(2003)}]{D'Elia:2002gd}%
  \BibitemOpen
  \bibfield  {author} {\bibinfo {author} {\bibfnamefont {M.}~\bibnamefont {D'Elia}}\ and\ \bibinfo {author} {\bibfnamefont {M.-P.}\ \bibnamefont {Lombardo}},\ }\href {\doibase 10.1103/PhysRevD.67.014505} {\bibfield  {journal} {\bibinfo  {journal} {Phys.Rev.}\ }\textbf {\bibinfo {volume} {D67}},\ \bibinfo {pages} {014505} (\bibinfo {year} {2003})},\ \Eprint {http://arxiv.org/abs/hep-lat/0209146} {arXiv:hep-lat/0209146 [hep-lat]} \BibitemShut {NoStop}%
\bibitem [{\citenamefont {D'Elia}\ and\ \citenamefont {Lombardo}(2004)}]{D'Elia:2004at}%
  \BibitemOpen
  \bibfield  {author} {\bibinfo {author} {\bibfnamefont {M.}~\bibnamefont {D'Elia}}\ and\ \bibinfo {author} {\bibfnamefont {M.-P.}\ \bibnamefont {Lombardo}},\ }\href {\doibase 10.1103/PhysRevD.70.074509} {\bibfield  {journal} {\bibinfo  {journal} {Phys.Rev.}\ }\textbf {\bibinfo {volume} {D70}},\ \bibinfo {pages} {074509} (\bibinfo {year} {2004})},\ \Eprint {http://arxiv.org/abs/hep-lat/0406012} {arXiv:hep-lat/0406012 [hep-lat]} \BibitemShut {NoStop}%
\bibitem [{\citenamefont {Chen}\ and\ \citenamefont {Luo}(2005)}]{Chen:2004tb}%
  \BibitemOpen
  \bibfield  {author} {\bibinfo {author} {\bibfnamefont {H.-S.}\ \bibnamefont {Chen}}\ and\ \bibinfo {author} {\bibfnamefont {X.-Q.}\ \bibnamefont {Luo}},\ }\href {\doibase 10.1103/PhysRevD.72.034504} {\bibfield  {journal} {\bibinfo  {journal} {Phys.Rev.}\ }\textbf {\bibinfo {volume} {D72}},\ \bibinfo {pages} {034504} (\bibinfo {year} {2005})},\ \Eprint {http://arxiv.org/abs/hep-lat/0411023} {arXiv:hep-lat/0411023 [hep-lat]} \BibitemShut {NoStop}%
\bibitem [{\citenamefont {D'Elia}\ and\ \citenamefont {Sanfilippo}(2009)}]{D'Elia:2009qz}%
  \BibitemOpen
  \bibfield  {author} {\bibinfo {author} {\bibfnamefont {M.}~\bibnamefont {D'Elia}}\ and\ \bibinfo {author} {\bibfnamefont {F.}~\bibnamefont {Sanfilippo}},\ }\href {\doibase 10.1103/PhysRevD.80.111501} {\bibfield  {journal} {\bibinfo  {journal} {Phys. Rev.}\ }\textbf {\bibinfo {volume} {D80}},\ \bibinfo {pages} {111501} (\bibinfo {year} {2009})},\ \Eprint {http://arxiv.org/abs/0909.0254} {arXiv:0909.0254 [hep-lat]} \BibitemShut {NoStop}%
\bibitem [{\citenamefont {Sakai}\ \emph {et~al.}(2009)\citenamefont {Sakai}, \citenamefont {Kashiwa}, \citenamefont {Kouno}, \citenamefont {Matsuzaki},\ and\ \citenamefont {Yahiro}}]{Sakai:2009dv}%
  \BibitemOpen
  \bibfield  {author} {\bibinfo {author} {\bibfnamefont {Y.}~\bibnamefont {Sakai}}, \bibinfo {author} {\bibfnamefont {K.}~\bibnamefont {Kashiwa}}, \bibinfo {author} {\bibfnamefont {H.}~\bibnamefont {Kouno}}, \bibinfo {author} {\bibfnamefont {M.}~\bibnamefont {Matsuzaki}}, \ and\ \bibinfo {author} {\bibfnamefont {M.}~\bibnamefont {Yahiro}},\ }\href {\doibase 10.1103/PhysRevD.79.096001} {\bibfield  {journal} {\bibinfo  {journal} {Phys. Rev. D}\ }\textbf {\bibinfo {volume} {79}},\ \bibinfo {pages} {096001} (\bibinfo {year} {2009})},\ \Eprint {http://arxiv.org/abs/0902.0487} {arXiv:0902.0487 [hep-ph]} \BibitemShut {NoStop}%
\bibitem [{\citenamefont {Cleymans}\ \emph {et~al.}(1986{\natexlab{a}})\citenamefont {Cleymans}, \citenamefont {Gavai},\ and\ \citenamefont {Suhonen}}]{Cleymans:1985wb}%
  \BibitemOpen
  \bibfield  {author} {\bibinfo {author} {\bibfnamefont {J.}~\bibnamefont {Cleymans}}, \bibinfo {author} {\bibfnamefont {R.~V.}\ \bibnamefont {Gavai}}, \ and\ \bibinfo {author} {\bibfnamefont {E.}~\bibnamefont {Suhonen}},\ }\href {\doibase 10.1016/0370-1573(86)90169-9} {\bibfield  {journal} {\bibinfo  {journal} {Phys. Rept.}\ }\textbf {\bibinfo {volume} {130}},\ \bibinfo {pages} {217} (\bibinfo {year} {1986}{\natexlab{a}})}\BibitemShut {NoStop}%
\bibitem [{\citenamefont {Cleymans}\ \emph {et~al.}(1986{\natexlab{b}})\citenamefont {Cleymans}, \citenamefont {Redlich}, \citenamefont {Satz},\ and\ \citenamefont {Suhonen}}]{Cleymans:1986cq}%
  \BibitemOpen
  \bibfield  {author} {\bibinfo {author} {\bibfnamefont {J.}~\bibnamefont {Cleymans}}, \bibinfo {author} {\bibfnamefont {K.}~\bibnamefont {Redlich}}, \bibinfo {author} {\bibfnamefont {H.}~\bibnamefont {Satz}}, \ and\ \bibinfo {author} {\bibfnamefont {E.}~\bibnamefont {Suhonen}},\ }\href {\doibase 10.1007/BF01410462} {\bibfield  {journal} {\bibinfo  {journal} {Z. Phys. C}\ }\textbf {\bibinfo {volume} {33}},\ \bibinfo {pages} {151} (\bibinfo {year} {1986}{\natexlab{b}})}\BibitemShut {NoStop}%
\bibitem [{\citenamefont {Kouno}\ and\ \citenamefont {Takagi}(1989)}]{Kouno:1988bi}%
  \BibitemOpen
  \bibfield  {author} {\bibinfo {author} {\bibfnamefont {H.}~\bibnamefont {Kouno}}\ and\ \bibinfo {author} {\bibfnamefont {F.}~\bibnamefont {Takagi}},\ }\href {\doibase 10.1007/BF01555858} {\bibfield  {journal} {\bibinfo  {journal} {Z. Phys. C}\ }\textbf {\bibinfo {volume} {42}},\ \bibinfo {pages} {209} (\bibinfo {year} {1989})}\BibitemShut {NoStop}%
\bibitem [{\citenamefont {Rischke}\ \emph {et~al.}(1991)\citenamefont {Rischke}, \citenamefont {Gorenstein}, \citenamefont {Stoecker},\ and\ \citenamefont {Greiner}}]{Rischke:1991ke}%
  \BibitemOpen
  \bibfield  {author} {\bibinfo {author} {\bibfnamefont {D.~H.}\ \bibnamefont {Rischke}}, \bibinfo {author} {\bibfnamefont {M.~I.}\ \bibnamefont {Gorenstein}}, \bibinfo {author} {\bibfnamefont {H.}~\bibnamefont {Stoecker}}, \ and\ \bibinfo {author} {\bibfnamefont {W.}~\bibnamefont {Greiner}},\ }\href {\doibase 10.1007/BF01548574} {\bibfield  {journal} {\bibinfo  {journal} {Z. Phys. C}\ }\textbf {\bibinfo {volume} {51}},\ \bibinfo {pages} {485} (\bibinfo {year} {1991})}\BibitemShut {NoStop}%
\bibitem [{\citenamefont {Miyahara}\ \emph {et~al.}(2020)\citenamefont {Miyahara}, \citenamefont {Ishii}, \citenamefont {Kouno},\ and\ \citenamefont {Yahiro}}]{Miyahara:2019zfn}%
  \BibitemOpen
  \bibfield  {author} {\bibinfo {author} {\bibfnamefont {A.}~\bibnamefont {Miyahara}}, \bibinfo {author} {\bibfnamefont {M.}~\bibnamefont {Ishii}}, \bibinfo {author} {\bibfnamefont {H.}~\bibnamefont {Kouno}}, \ and\ \bibinfo {author} {\bibfnamefont {M.}~\bibnamefont {Yahiro}},\ }\href {\doibase 10.1103/PhysRevD.101.076011} {\bibfield  {journal} {\bibinfo  {journal} {Phys. Rev. D}\ }\textbf {\bibinfo {volume} {101}},\ \bibinfo {pages} {076011} (\bibinfo {year} {2020})},\ \Eprint {http://arxiv.org/abs/1907.07306} {arXiv:1907.07306 [hep-ph]} \BibitemShut {NoStop}%
\bibitem [{\citenamefont {Jeong}\ \emph {et~al.}(2020)\citenamefont {Jeong}, \citenamefont {McLerran},\ and\ \citenamefont {Sen}}]{Jeong:2019lhv}%
  \BibitemOpen
  \bibfield  {author} {\bibinfo {author} {\bibfnamefont {K.~S.}\ \bibnamefont {Jeong}}, \bibinfo {author} {\bibfnamefont {L.}~\bibnamefont {McLerran}}, \ and\ \bibinfo {author} {\bibfnamefont {S.}~\bibnamefont {Sen}},\ }\href {\doibase 10.1103/PhysRevC.101.035201} {\bibfield  {journal} {\bibinfo  {journal} {Phys. Rev. C}\ }\textbf {\bibinfo {volume} {101}},\ \bibinfo {pages} {035201} (\bibinfo {year} {2020})},\ \Eprint {http://arxiv.org/abs/1908.04799} {arXiv:1908.04799 [nucl-th]} \BibitemShut {NoStop}%
\bibitem [{\citenamefont {Fujimoto}\ \emph {et~al.}(2022)\citenamefont {Fujimoto}, \citenamefont {Fukushima}, \citenamefont {Hidaka}, \citenamefont {Hiraguchi},\ and\ \citenamefont {Iida}}]{Fujimoto:2021dvn}%
  \BibitemOpen
  \bibfield  {author} {\bibinfo {author} {\bibfnamefont {Y.}~\bibnamefont {Fujimoto}}, \bibinfo {author} {\bibfnamefont {K.}~\bibnamefont {Fukushima}}, \bibinfo {author} {\bibfnamefont {Y.}~\bibnamefont {Hidaka}}, \bibinfo {author} {\bibfnamefont {A.}~\bibnamefont {Hiraguchi}}, \ and\ \bibinfo {author} {\bibfnamefont {K.}~\bibnamefont {Iida}},\ }\href {\doibase 10.1016/j.physletb.2022.137524} {\bibfield  {journal} {\bibinfo  {journal} {Phys. Lett. B}\ }\textbf {\bibinfo {volume} {835}},\ \bibinfo {pages} {137524} (\bibinfo {year} {2022})},\ \Eprint {http://arxiv.org/abs/2109.06799} {arXiv:2109.06799 [nucl-th]} \BibitemShut {NoStop}%
\bibitem [{\citenamefont {Roberge}\ and\ \citenamefont {Weiss}(1986)}]{Roberge:1986mm}%
  \BibitemOpen
  \bibfield  {author} {\bibinfo {author} {\bibfnamefont {A.}~\bibnamefont {Roberge}}\ and\ \bibinfo {author} {\bibfnamefont {N.}~\bibnamefont {Weiss}},\ }\href {\doibase 10.1016/0550-3213(86)90582-1} {\bibfield  {journal} {\bibinfo  {journal} {Nucl. Phys. B}\ }\textbf {\bibinfo {volume} {275}},\ \bibinfo {pages} {734} (\bibinfo {year} {1986})}\BibitemShut {NoStop}%
\bibitem [{\citenamefont {Bonati}\ \emph {et~al.}(2016)\citenamefont {Bonati}, \citenamefont {D'Elia}, \citenamefont {Mariti}, \citenamefont {Mesiti}, \citenamefont {Negro},\ and\ \citenamefont {Sanfilippo}}]{Bonati:2016pwz}%
  \BibitemOpen
  \bibfield  {author} {\bibinfo {author} {\bibfnamefont {C.}~\bibnamefont {Bonati}}, \bibinfo {author} {\bibfnamefont {M.}~\bibnamefont {D'Elia}}, \bibinfo {author} {\bibfnamefont {M.}~\bibnamefont {Mariti}}, \bibinfo {author} {\bibfnamefont {M.}~\bibnamefont {Mesiti}}, \bibinfo {author} {\bibfnamefont {F.}~\bibnamefont {Negro}}, \ and\ \bibinfo {author} {\bibfnamefont {F.}~\bibnamefont {Sanfilippo}},\ }\href {\doibase 10.1103/PhysRevD.93.074504} {\bibfield  {journal} {\bibinfo  {journal} {Phys. Rev. D}\ }\textbf {\bibinfo {volume} {93}},\ \bibinfo {pages} {074504} (\bibinfo {year} {2016})},\ \Eprint {http://arxiv.org/abs/1602.01426} {arXiv:1602.01426 [hep-lat]} \BibitemShut {NoStop}%
\bibitem [{\citenamefont {Cuteri}\ \emph {et~al.}(2022)\citenamefont {Cuteri}, \citenamefont {Goswami}, \citenamefont {Karsch}, \citenamefont {Lahiri}, \citenamefont {Neumann}, \citenamefont {Philipsen}, \citenamefont {Schmidt},\ and\ \citenamefont {Sciarra}}]{Cuteri:2022vwk}%
  \BibitemOpen
  \bibfield  {author} {\bibinfo {author} {\bibfnamefont {F.}~\bibnamefont {Cuteri}}, \bibinfo {author} {\bibfnamefont {J.}~\bibnamefont {Goswami}}, \bibinfo {author} {\bibfnamefont {F.}~\bibnamefont {Karsch}}, \bibinfo {author} {\bibfnamefont {A.}~\bibnamefont {Lahiri}}, \bibinfo {author} {\bibfnamefont {M.}~\bibnamefont {Neumann}}, \bibinfo {author} {\bibfnamefont {O.}~\bibnamefont {Philipsen}}, \bibinfo {author} {\bibfnamefont {C.}~\bibnamefont {Schmidt}}, \ and\ \bibinfo {author} {\bibfnamefont {A.}~\bibnamefont {Sciarra}},\ }\href {\doibase 10.1103/PhysRevD.106.014510} {\bibfield  {journal} {\bibinfo  {journal} {Phys. Rev. D}\ }\textbf {\bibinfo {volume} {106}},\ \bibinfo {pages} {014510} (\bibinfo {year} {2022})},\ \Eprint {http://arxiv.org/abs/2205.12707} {arXiv:2205.12707 [hep-lat]} \BibitemShut {NoStop}%
\bibitem [{\citenamefont {Bonati}\ \emph {et~al.}(2019)\citenamefont {Bonati}, \citenamefont {Calore}, \citenamefont {D'Elia}, \citenamefont {Mesiti}, \citenamefont {Negro}, \citenamefont {Sanfilippo}, \citenamefont {Schifano}, \citenamefont {Silvi},\ and\ \citenamefont {Tripiccione}}]{Bonati:2018fvg}%
  \BibitemOpen
  \bibfield  {author} {\bibinfo {author} {\bibfnamefont {C.}~\bibnamefont {Bonati}}, \bibinfo {author} {\bibfnamefont {E.}~\bibnamefont {Calore}}, \bibinfo {author} {\bibfnamefont {M.}~\bibnamefont {D'Elia}}, \bibinfo {author} {\bibfnamefont {M.}~\bibnamefont {Mesiti}}, \bibinfo {author} {\bibfnamefont {F.}~\bibnamefont {Negro}}, \bibinfo {author} {\bibfnamefont {F.}~\bibnamefont {Sanfilippo}}, \bibinfo {author} {\bibfnamefont {S.~F.}\ \bibnamefont {Schifano}}, \bibinfo {author} {\bibfnamefont {G.}~\bibnamefont {Silvi}}, \ and\ \bibinfo {author} {\bibfnamefont {R.}~\bibnamefont {Tripiccione}},\ }\href {\doibase 10.1103/PhysRevD.99.014502} {\bibfield  {journal} {\bibinfo  {journal} {Phys. Rev. D}\ }\textbf {\bibinfo {volume} {99}},\ \bibinfo {pages} {014502} (\bibinfo {year} {2019})},\ \Eprint {http://arxiv.org/abs/1807.02106} {arXiv:1807.02106 [hep-lat]} \BibitemShut {NoStop}%
\bibitem [{\citenamefont {Taradiy}\ \emph {et~al.}(2019)\citenamefont {Taradiy}, \citenamefont {Motornenko}, \citenamefont {Vovchenko}, \citenamefont {Gorenstein},\ and\ \citenamefont {Stoecker}}]{Taradiy:2019taz}%
  \BibitemOpen
  \bibfield  {author} {\bibinfo {author} {\bibfnamefont {K.}~\bibnamefont {Taradiy}}, \bibinfo {author} {\bibfnamefont {A.}~\bibnamefont {Motornenko}}, \bibinfo {author} {\bibfnamefont {V.}~\bibnamefont {Vovchenko}}, \bibinfo {author} {\bibfnamefont {M.~I.}\ \bibnamefont {Gorenstein}}, \ and\ \bibinfo {author} {\bibfnamefont {H.}~\bibnamefont {Stoecker}},\ }\href {\doibase 10.1103/PhysRevC.100.065202} {\bibfield  {journal} {\bibinfo  {journal} {Phys. Rev. C}\ }\textbf {\bibinfo {volume} {100}},\ \bibinfo {pages} {065202} (\bibinfo {year} {2019})},\ \Eprint {http://arxiv.org/abs/1904.08259} {arXiv:1904.08259 [hep-ph]} \BibitemShut {NoStop}%
\bibitem [{\citenamefont {Savchuk}\ \emph {et~al.}(2020)\citenamefont {Savchuk}, \citenamefont {Vovchenko}, \citenamefont {Poberezhnyuk}, \citenamefont {Gorenstein},\ and\ \citenamefont {Stoecker}}]{Savchuk:2019yxl}%
  \BibitemOpen
  \bibfield  {author} {\bibinfo {author} {\bibfnamefont {O.}~\bibnamefont {Savchuk}}, \bibinfo {author} {\bibfnamefont {V.}~\bibnamefont {Vovchenko}}, \bibinfo {author} {\bibfnamefont {R.~V.}\ \bibnamefont {Poberezhnyuk}}, \bibinfo {author} {\bibfnamefont {M.~I.}\ \bibnamefont {Gorenstein}}, \ and\ \bibinfo {author} {\bibfnamefont {H.}~\bibnamefont {Stoecker}},\ }\href {\doibase 10.1103/PhysRevC.101.035205} {\bibfield  {journal} {\bibinfo  {journal} {Phys. Rev. C}\ }\textbf {\bibinfo {volume} {101}},\ \bibinfo {pages} {035205} (\bibinfo {year} {2020})},\ \Eprint {http://arxiv.org/abs/1909.04461} {arXiv:1909.04461 [hep-ph]} \BibitemShut {NoStop}%
\bibitem [{\citenamefont {Fukushima}(2004)}]{Fukushima:2003fw}%
  \BibitemOpen
  \bibfield  {author} {\bibinfo {author} {\bibfnamefont {K.}~\bibnamefont {Fukushima}},\ }\href {\doibase 10.1016/j.physletb.2004.04.027} {\bibfield  {journal} {\bibinfo  {journal} {Phys. Lett. B}\ }\textbf {\bibinfo {volume} {591}},\ \bibinfo {pages} {277} (\bibinfo {year} {2004})},\ \Eprint {http://arxiv.org/abs/hep-ph/0310121} {arXiv:hep-ph/0310121} \BibitemShut {NoStop}%
\bibitem [{\citenamefont {Ratti}\ \emph {et~al.}(2006)\citenamefont {Ratti}, \citenamefont {Thaler},\ and\ \citenamefont {Weise}}]{Ratti:2005jh}%
  \BibitemOpen
  \bibfield  {author} {\bibinfo {author} {\bibfnamefont {C.}~\bibnamefont {Ratti}}, \bibinfo {author} {\bibfnamefont {M.~A.}\ \bibnamefont {Thaler}}, \ and\ \bibinfo {author} {\bibfnamefont {W.}~\bibnamefont {Weise}},\ }\href {\doibase 10.1103/PhysRevD.73.014019} {\bibfield  {journal} {\bibinfo  {journal} {Phys. Rev. D}\ }\textbf {\bibinfo {volume} {73}},\ \bibinfo {pages} {014019} (\bibinfo {year} {2006})},\ \Eprint {http://arxiv.org/abs/hep-ph/0506234} {arXiv:hep-ph/0506234} \BibitemShut {NoStop}%
\bibitem [{\citenamefont {Ghosh}\ \emph {et~al.}(2006)\citenamefont {Ghosh}, \citenamefont {Mukherjee}, \citenamefont {Mustafa},\ and\ \citenamefont {Ray}}]{Ghosh:2006qh}%
  \BibitemOpen
  \bibfield  {author} {\bibinfo {author} {\bibfnamefont {S.~K.}\ \bibnamefont {Ghosh}}, \bibinfo {author} {\bibfnamefont {T.~K.}\ \bibnamefont {Mukherjee}}, \bibinfo {author} {\bibfnamefont {M.~G.}\ \bibnamefont {Mustafa}}, \ and\ \bibinfo {author} {\bibfnamefont {R.}~\bibnamefont {Ray}},\ }\href {\doibase 10.1103/PhysRevD.73.114007} {\bibfield  {journal} {\bibinfo  {journal} {Phys. Rev. D}\ }\textbf {\bibinfo {volume} {73}},\ \bibinfo {pages} {114007} (\bibinfo {year} {2006})},\ \Eprint {http://arxiv.org/abs/hep-ph/0603050} {arXiv:hep-ph/0603050} \BibitemShut {NoStop}%
\bibitem [{\citenamefont {Megias}\ \emph {et~al.}(2006)\citenamefont {Megias}, \citenamefont {Ruiz~Arriola},\ and\ \citenamefont {Salcedo}}]{Megias:2004hj}%
  \BibitemOpen
  \bibfield  {author} {\bibinfo {author} {\bibfnamefont {E.}~\bibnamefont {Megias}}, \bibinfo {author} {\bibfnamefont {E.}~\bibnamefont {Ruiz~Arriola}}, \ and\ \bibinfo {author} {\bibfnamefont {L.~L.}\ \bibnamefont {Salcedo}},\ }\href {\doibase 10.1103/PhysRevD.74.065005} {\bibfield  {journal} {\bibinfo  {journal} {Phys. Rev. D}\ }\textbf {\bibinfo {volume} {74}},\ \bibinfo {pages} {065005} (\bibinfo {year} {2006})},\ \Eprint {http://arxiv.org/abs/hep-ph/0412308} {arXiv:hep-ph/0412308} \BibitemShut {NoStop}%
\bibitem [{\citenamefont {R{\"{o}}{\ss}ner}\ \emph {et~al.}(2007)\citenamefont {R{\"{o}}{\ss}ner}, \citenamefont {Ratti},\ and\ \citenamefont {Weise}}]{Roessner:2006xn}%
  \BibitemOpen
  \bibfield  {author} {\bibinfo {author} {\bibfnamefont {S.}~\bibnamefont {R{\"{o}}{\ss}ner}}, \bibinfo {author} {\bibfnamefont {C.}~\bibnamefont {Ratti}}, \ and\ \bibinfo {author} {\bibfnamefont {W.}~\bibnamefont {Weise}},\ }\href {\doibase 10.1103/PhysRevD.75.034007} {\bibfield  {journal} {\bibinfo  {journal} {Phys. Rev. D}\ }\textbf {\bibinfo {volume} {75}},\ \bibinfo {pages} {034007} (\bibinfo {year} {2007})},\ \Eprint {http://arxiv.org/abs/hep-ph/0609281} {arXiv:hep-ph/0609281} \BibitemShut {NoStop}%
\bibitem [{\citenamefont {Sakai}\ \emph {et~al.}(2010)\citenamefont {Sakai}, \citenamefont {Sasaki}, \citenamefont {Kouno},\ and\ \citenamefont {Yahiro}}]{Sakai:2010rp}%
  \BibitemOpen
  \bibfield  {author} {\bibinfo {author} {\bibfnamefont {Y.}~\bibnamefont {Sakai}}, \bibinfo {author} {\bibfnamefont {T.}~\bibnamefont {Sasaki}}, \bibinfo {author} {\bibfnamefont {H.}~\bibnamefont {Kouno}}, \ and\ \bibinfo {author} {\bibfnamefont {M.}~\bibnamefont {Yahiro}},\ }\href {\doibase 10.1103/PhysRevD.82.076003} {\bibfield  {journal} {\bibinfo  {journal} {Phys. Rev. D}\ }\textbf {\bibinfo {volume} {82}},\ \bibinfo {pages} {076003} (\bibinfo {year} {2010})},\ \Eprint {http://arxiv.org/abs/1006.3648} {arXiv:1006.3648 [hep-ph]} \BibitemShut {NoStop}%
\bibitem [{\citenamefont {Kashiwa}(2019)}]{Kashiwa:2019ihm}%
  \BibitemOpen
  \bibfield  {author} {\bibinfo {author} {\bibfnamefont {K.}~\bibnamefont {Kashiwa}},\ }\href {\doibase 10.3390/sym11040562} {\bibfield  {journal} {\bibinfo  {journal} {Symmetry}\ }\textbf {\bibinfo {volume} {11}},\ \bibinfo {pages} {562} (\bibinfo {year} {2019})}\BibitemShut {NoStop}%
\bibitem [{\citenamefont {Workman}\ and\ \citenamefont {Others}(2022)}]{Workman:2022ynf}%
  \BibitemOpen
  \bibfield  {author} {\bibinfo {author} {\bibfnamefont {R.~L.}\ \bibnamefont {Workman}}\ and\ \bibinfo {author} {\bibnamefont {Others}} (\bibinfo {collaboration} {Particle Data Group}),\ }\href {\doibase 10.1093/ptep/ptac097} {\bibfield  {journal} {\bibinfo  {journal} {PTEP}\ }\textbf {\bibinfo {volume} {2022}},\ \bibinfo {pages} {083C01} (\bibinfo {year} {2022})}\BibitemShut {NoStop}%
\bibitem [{\citenamefont {Vovchenko}\ \emph {et~al.}(2017)\citenamefont {Vovchenko}, \citenamefont {Pasztor}, \citenamefont {Fodor}, \citenamefont {Katz},\ and\ \citenamefont {Stoecker}}]{Vovchenko:2017xad}%
  \BibitemOpen
  \bibfield  {author} {\bibinfo {author} {\bibfnamefont {V.}~\bibnamefont {Vovchenko}}, \bibinfo {author} {\bibfnamefont {A.}~\bibnamefont {Pasztor}}, \bibinfo {author} {\bibfnamefont {Z.}~\bibnamefont {Fodor}}, \bibinfo {author} {\bibfnamefont {S.~D.}\ \bibnamefont {Katz}}, \ and\ \bibinfo {author} {\bibfnamefont {H.}~\bibnamefont {Stoecker}},\ }\href {\doibase 10.1016/j.physletb.2017.10.042} {\bibfield  {journal} {\bibinfo  {journal} {Phys. Lett. B}\ }\textbf {\bibinfo {volume} {775}},\ \bibinfo {pages} {71} (\bibinfo {year} {2017})},\ \Eprint {http://arxiv.org/abs/1708.02852} {arXiv:1708.02852 [hep-ph]} \BibitemShut {NoStop}%
\bibitem [{\citenamefont {Kouno}\ and\ \citenamefont {Kashiwa}(2024)}]{Kouno:2023ygw}%
  \BibitemOpen
  \bibfield  {author} {\bibinfo {author} {\bibfnamefont {H.}~\bibnamefont {Kouno}}\ and\ \bibinfo {author} {\bibfnamefont {K.}~\bibnamefont {Kashiwa}},\ }\href {\doibase 10.1103/PhysRevD.109.054007} {\bibfield  {journal} {\bibinfo  {journal} {Phys. Rev. D}\ }\textbf {\bibinfo {volume} {109}},\ \bibinfo {pages} {054007} (\bibinfo {year} {2024})},\ \Eprint {http://arxiv.org/abs/2310.09738} {arXiv:2310.09738 [hep-ph]} \BibitemShut {NoStop}%
\bibitem [{\citenamefont {Rehberg}\ \emph {et~al.}(1996)\citenamefont {Rehberg}, \citenamefont {Klevansky},\ and\ \citenamefont {Hufner}}]{Rehberg:1995kh}%
  \BibitemOpen
  \bibfield  {author} {\bibinfo {author} {\bibfnamefont {P.}~\bibnamefont {Rehberg}}, \bibinfo {author} {\bibfnamefont {S.~P.}\ \bibnamefont {Klevansky}}, \ and\ \bibinfo {author} {\bibfnamefont {J.}~\bibnamefont {Hufner}},\ }\href {\doibase 10.1103/PhysRevC.53.410} {\bibfield  {journal} {\bibinfo  {journal} {Phys. Rev. C}\ }\textbf {\bibinfo {volume} {53}},\ \bibinfo {pages} {410} (\bibinfo {year} {1996})},\ \Eprint {http://arxiv.org/abs/hep-ph/9506436} {arXiv:hep-ph/9506436} \BibitemShut {NoStop}%
\bibitem [{\citenamefont {Miyahara}\ \emph {et~al.}(2017)\citenamefont {Miyahara}, \citenamefont {Ishii}, \citenamefont {Kouno},\ and\ \citenamefont {Yahiro}}]{Miyahara:2017eam}%
  \BibitemOpen
  \bibfield  {author} {\bibinfo {author} {\bibfnamefont {A.}~\bibnamefont {Miyahara}}, \bibinfo {author} {\bibfnamefont {M.}~\bibnamefont {Ishii}}, \bibinfo {author} {\bibfnamefont {H.}~\bibnamefont {Kouno}}, \ and\ \bibinfo {author} {\bibfnamefont {M.}~\bibnamefont {Yahiro}},\ }\href {\doibase 10.1142/S0217751X17502050} {\bibfield  {journal} {\bibinfo  {journal} {Int. J. Mod. Phys. A}\ }\textbf {\bibinfo {volume} {32}},\ \bibinfo {pages} {1750205} (\bibinfo {year} {2017})},\ \Eprint {http://arxiv.org/abs/1704.06432} {arXiv:1704.06432 [hep-ph]} \BibitemShut {NoStop}%
\end{thebibliography}%

\end{document}